\documentclass[lettersize,journal]{IEEEtran}
\usepackage{amsmath,amsfonts}
\usepackage{algorithm}
\usepackage{array}
\usepackage[caption=false,font=normalsize,labelfont=sf,textfont=sf]{subfig}
\usepackage{textcomp}
\usepackage{stfloats}
\usepackage{url}
\usepackage{verbatim}
\usepackage{graphicx}
\usepackage{cite}
\newcommand{\ewu}[1]{\textcolor{red}{\textbf{ewu: #1\xspace}}}

\newcommand{\yilan}[1]{\textcolor{orange}{\textbf{yilan: #1\xspace}}}

\newcommand{\cindy}[1]{\textcolor{blue}{\textbf{cindy: #1\xspace}}}

\DeclareMathOperator*{\argmax}{arg\,max}

\newcommand{\eat}[1]{}

\newcommand{\stitle}[1]{\vspace{2pt}\noindent\textbf{#1}}

\usepackage{ifpdf}

\ifpdf
  \pdfoutput=1\relax                   
  \usepackage{graphicx}                
  \DeclareGraphicsExtensions{.pdf,.png,.jpg,.jpeg} 
\else
  \usepackage{graphicx}                
  \DeclareGraphicsExtensions{.eps}     
\fi%

\graphicspath{{figures/}{pictures/}{images/}{./}} 

\usepackage{microtype}                 
\PassOptionsToPackage{warn}{textcomp}  
\usepackage{textcomp}                  
\usepackage{mathptmx}                  
\usepackage{times}                     
\usepackage{cite}                      
\usepackage{tabu}                      
\usepackage{booktabs}                  
\usepackage{enumitem}
\usepackage{listings}
\usepackage{algorithm}
\usepackage{algpseudocode}
\usepackage{color}
\usepackage{amsmath}
\usepackage{amssymb}
 \usepackage{amsthm}

\usepackage{tabularx}

\usepackage{makecell}
\usepackage[table]{xcolor}

\usepackage{xcolor}

\usepackage{framed,lipsum}

\usepackage[T1]{fontenc}
\usepackage[scaled=0.88]{beramono}    
\usepackage{listings}
\newtheorem{example}{Example}

\newcounter{prob}
\newtheorem{problem}[prob]{Problem}

\newlength{\listingindent}                
\lstdefinelanguage{goh}
{morekeywords={var,hyp,attr,const,expr,func,pred,op,except,
               expr1,expr2,var1,var2,attr1,attr2,const1,const2,const3,num1,num2,hyp1,hyp2},
sensitive=false,
morecomment=[l]{//},
morecomment=[s]{/*}{*/},
morestring=[b]",
}

\definecolor{light-gray}{gray}{0.95}
\definecolor{mid-gray}{gray}{0.65}
\definecolor{darkred}{rgb}{0.7,0.25,0.25}
\definecolor{darkgreen}{rgb}{0.15,0.55,0.15}
\definecolor{darkblue}{rgb}{0.1,0.1,0.5}
\definecolor{blue}{rgb}{0.19,0.58,1}

\newcommand{\gray}[1]{\textcolor{mid-gray}{#1}}
\newcommand{\red}[1]{\textcolor{red}{#1}}

\newcommand{\blue}[1]{\textcolor{blue}{#1}}

\definecolor{salmon}{RGB}{232, 125, 114}

\definecolor{mred}{rgb}{.80,.12,.30}
\definecolor{grey}{rgb}{.5,.5,.5}
\definecolor{turquoise}{rgb}{.04,.79,.93}

\newif\ifnotes
\notestrue

\usepackage[normalem]{ulem}
\usepackage{soul}
\usepackage{hyperref}
\usepackage[noabbrev,capitalise]{cleveref}

\let\origcite\cite
\renewcommand{\cite}[1]{\ifnotes\mbox{\origcite{#1}}\else \origcite{#1}\fi}

\lstdefinestyle{SQLStyle}{
  language=SQL,
  showspaces=false,
  basicstyle=\ttfamily\scriptsize,
  commentstyle=\color{gray},
  mathescape=true,
  numbers=none,
  escapeinside={^}{^},
  captionpos=b,
  float=tp,
  floatplacement=tbp,
  belowskip=-0.05em,
   mathescape=false
}

\newlength{\leftbarwidth}
\newlength{\leftbarsep}
\colorlet{leftbarcolor}{gray}

\def\yilan#1{{{\color{black} #1}}}
\def\cindy#1{{{\color{black} #1}}}

\begin{document}

\title{Data-Induced Groupings and How To Find Them}

\author{Yilan Jiang, Cindy Xiong Bearfield, Steven Franconeri, Eugene Wu
\thanks{Yilan Jiang is with the University of Illinois Urbana-Champaign, Urbana, IL 61801 USA
(e-mail: yilanj2@illinois.edu).}
\thanks{Cindy Xiong Bearfield is with Georgia Institute of Technology, Atlanta, GA 30332 USA
(e-mail: cxiong@gatech.edu).}
\thanks{Steven Franconeri is with Northwestern University,
Evanston, IL 60208 USA
(e-mail: franconeri@northwestern.edu).}
\thanks{Eugene Wu is with Columbia University,
New York, NY 10027 USA
(e-mail: ewu@cs.columbia.edu).}

}



\maketitle

\begin{abstract}
Making sense of a visualization requires the reader to consider both the visualization design and the underlying data values. 
Existing work in the visualization community has largely considered affordances driven by visualization design elements, such as color or chart type, but how visual design interacts with data values to impact interpretation and reasoning has remained under-explored. 
Dot plots and bar graphs are commonly used to help users identify groups of points that form trends and clusters, but are liable to manifest groupings that are artifacts of spatial arrangement rather than inherent patterns in the data itself. 
These ``Data-induced Groups'' can drive suboptimal data comparisons and potentially lead the user to incorrect conclusions. 
We conduct two user studies using dot plots as a case study to understand the prevalence of data-induced groupings.
We find that users rely on data-induced groupings in both conditions despite the fact that trend-based groupings are irrelevant in nominal data. 
Based on the study results, we build a model to predict whether users are likely to perceive a given set of dot plot points as a group.
We discuss two use cases illustrating how the model can assist visualization designers by both diagnosing potential user-perceived groupings in dot plots and offering redesigns that better accentuate desired groupings through data rearrangement.
\end{abstract}


\section{Introduction}
\IEEEPARstart{E}{ffective} visualizations enable readers to extract key messages and compare data quickly, while poor designs can mislead or overwhelm.
Existing work in the visualization community has generated guidelines that help people choose the appropriate designs based on reader intentions (e.g., analytic goals), data type (e.g., categorical, continuous), or data distributions (e.g., total number of points, correlation between attributes) ~\cite{mackinlay1986automating, kim2018assessing, franconeri2021science, quadri2021survey}.
For example, scatterplots afford clustering and correlation tasks~\cite{saket2018task, yang2018correlation}, but are less ideal for summary tasks such as comparing the averages of two classes, especially as the number of data points increases~\cite{kim2018assessing}.
Line charts are helpful for trend detection with quantitative data and can even mitigate causation bias~\cite{best2007perception, xiong2019illusion}, but become misleading when used to visualize categorical data~\cite{zacks1999bars}.

We posit that visualization designers should also consider the interplay between design choices and the underlying \textit{data values}. 
Depending on the spatial arrangement of data values, for example, users may see groupings or trends that are artifacts of our perceptual encoding mechanisms and not present in the data.  
Regardless of user goals, these artificial groupings can cause users to see particular relationships or visual comparisons over others, and thus lead to invalid conclusions.  

\begin{figure}[h]
 \centering 
 \includegraphics[width = \columnwidth, trim={0cm 0.5cm 0cm 0cm}]{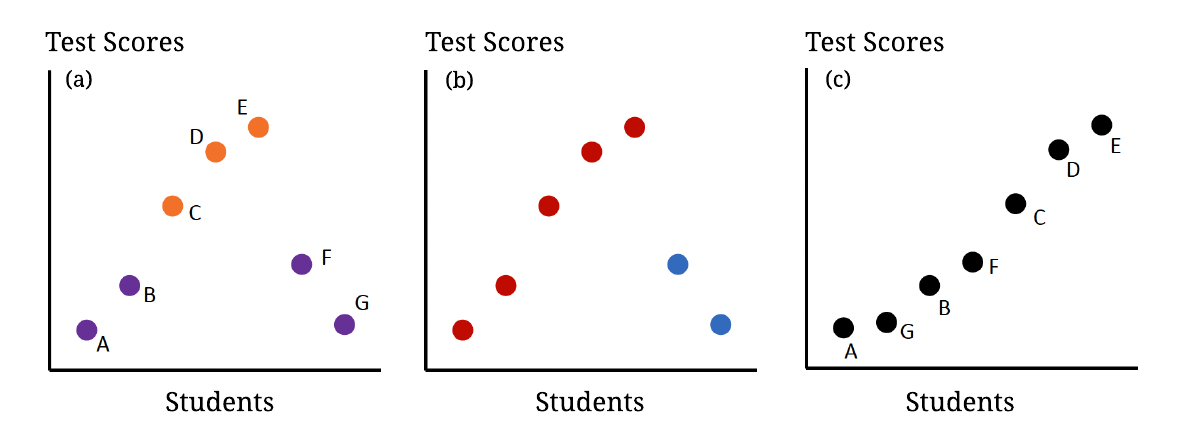}
 \caption{Nominal data showing test scores of seven students. One could see the three highest performing students as a group and the rest as another (a), or see a `growing trend' in the left five data points with the two points on the right as outliers (b). However, order along the nomial x axis is not meaningful. Rearranging the x-axis may encourage the viewer to see a `growing trend' across all seven points (c).}
 \label{hallucinator_groupings}
\end{figure}

In this paper, we aim to understand how users group data using dot plots as our study setting.  
Although bar charts are a conventional choice for summarizing data, a growing body of research has demonstrated that they introduce their own systematic biases, such as within-the-bar and bar-tip-limit effects~\cite{newman2012bar, kerns2021two}. 
As a result, recent work has increasingly advocated for dot-based representations, often accompanied by variability indicators, as a clearer and less biased alternative~\cite{holder2022dispersion}. 
This shift makes it important to examine grouping behaviors specifically in dot plots.
We created an example dot plot, as shown in \Cref{hallucinator_groupings}, which renders nominal data of seven student test scores, and different x-axis orderings accentuate different groupings and patterns.
A reader might group the top three highest-scoring students, or the left five data points, separately from the rest.
Alternatively, the reader might notice an `upward trend' across all students, which is misleading because the x-axis is nominal.

In this example, the perceived groupings change while the underlying data has remained the same. 
Scheidegger and Kinneman terms these groupings as {\it hallucinators}~\cite{kindlmann2014algebraic}  because the visual representation (and thus perceived patterns) changed while the data did not.  
These hallucinated groupings induce an implicit ``group id'' channel in the underlying data.
For instance, \Cref{hallucinator_groupings}(left) induces an implicit grouping attribute where CDE form group 1 and ABFG form group 2.   
We refer to these as {\it Data-induced Groupings}.

Data-induced groupings are related to perceptual grouping in the human perception literature. While similar to the concept of {\it hallucinators} which refers to unintended patterns perceived by users, data-induced groupings focus specifically on the influence of spatial configurations and excludes broader cognitive biases. 
During data interpretation, a reader dissects the visualization into units by visually grouping data points~\cite{franconeri2021science}.
This process allows the reader to make comparisons and compute data statistics needed to generate takeaways.  
Critically, once a reader has perceptually grouped certain data points together, it becomes extremely effortful for them to `break' those groups, form new groups, and compare those new groups~\cite{bearfield2023does, lovett2017modeling}. 
Therefore designers should thoughtfully build visualizations to facilitate the `right' grouping behavior.  
Existing work has demonstrated that visualization design, such as color and spatial arrangement choices, impact how readers visually group data and compare data~\cite{ fygenson2023arrangement, bearfield2023does}. 
However, the researchers have approached their user studies by randomizing the data values to account for their potential effects, therefore the effect of the underlying data values on grouping behaviors, and how data values might interact with design choices to impact viewer takeaways, have remained unexplored. 
Here, we build data-driven taxonomies and predictive models to advance our understanding of data-induced grouping behaviors in visualizations.
\cindy{Following prior visualization work that demonstrates that simple visualizations with as few as six data points can inform researchers of underlying perceptual grouping mechanisms when interpreting visualizations~\cite{xiong2021visual}, we adopt a similar design to examine dot plots with six data points.  
Studying these simpler cases provides a tractable way to uncover underlying perceptual mechanisms before extending to larger and more complex datasets~\cite{rensink2013prospects}.
This investigation allows us to offer an automated approach that reveals broader trends in human grouping behaviors so practitioners can account for the effect of data values in visualization design, mitigating the harmful effects of data-induced groupings by empowering designers to break-up hallucinated groups early on in the design process.
}

This paper makes three main contributions.
\begin{itemize}[leftmargin=*]
    \item  First, we conduct two user studies to understand the prevalence and characteristics of data-induced grouping channels and the capability for instructions to counteract the visual grouping effects. 
    Both studies render dot plots of 6 random points, where the x-axis is nominal and y-axis is numeric.   
    The first study asks participants to identify perceived groupings, while the second study additionally alerts participants of the nominal x-axis.  
    We study whether participants rely on trends or clustering based on the x-axis, even when they are explicitly cautioned about its nominal nature.
    \item Second, we use the results of the user studies to build predictive models that, given a set of points in a dot plot, predict the likelihood that a user will perceive the points as a group.  
    The models extract interpretable features based on co-linearity and clustering properties of the set.  On a hold-out dataset, our models achieve an F1 score (harmonic mean of precision and recall) of $97-99\%$.
    \item Third, we describe two applications of the model: as a diagnostic tool that predicts the data-induced groupings in a dot plot, and as a redesign tool to help visualization creators induce or break data-induced groupings. We integrate these functionalities in a visualization design tool that helps designers anticipate potential data-induced groupings, and search hundreds of possible redesigns to find designs that emphasize groups that aid their desired comparisons.
\end{itemize}

\section{Related Work}
Recognizing a visual configuration involves not only identifying individual components but also understanding the relationships among them. This process, known as perceptual organization, refers to recognizing how different elements within a scene relate to each other~\cite{wertheimer1938laws}. 
Through this mechanism, visual elements are grouped, and structure is imposed to form high-level visual objects.

Perceptual organization plays a critical role in how we interpret visualizations. 
While viewers can apply top-down control to influence how a visualization is organized, perceptual organization also naturally builds hierarchical visual representations from lower-level components. 
Theories of perceptual organization are largely based on the Gestalt Principles of grouping~\cite{wagemans2012century}. 
For example, the principle of proximity indicates that elements positioned close to each other are often grouped together, while the principle of similarity suggests that objects with similar features, such as colors or shapes, are perceived as a group~\cite{palmer1977hierarchical}.
In this work, we focus on the principle of continuity, which states that elements are grouped when they form a continuous contour~\cite{prinzmetal1977good}, such as aligned points in a dot plot.

\subsection{Comparisons in Visualization}

Comparison is a critical visual analysis task~\cite{gleicher2017considerations}.
For example, in visual analytics, prevalent user objectives involve comparing observed data to hypothesized results, or comparing patterns prior to and after a code-change or between manipulations \cite{rae2022understanding, battle2019characterizing}. 
When making comparisons, people always need to group the visual elements first, as grouping defines the units available for comparison. 
The visual system often prioritizes ``the forest before the trees,'' perceiving groups of objects more easily than individual ones \cite{navon1977forest,kimchi1992primacy}.
This happens because discerning individual objects require more time and attention~\cite{ahissar2004reverse}.
In the context of visualizations, past research suggests that spatial cues are among the most powerful grouping mechanisms~\cite{bearfield2023does}. 
As a result, we can predict that the easiest comparisons will involve data points that are spatially grouped, where proximity between points facilitates quicker and more intuitive comparisons.

The visualization research community has  developed tools to support visual comparisons of data. 
For example, \cite{wu2022view} proposed a conceptual model that allows people to summarize data and compute differences between data marks, trends, and other chart elements to facilitate ad hoc comparisons during data exploration.
Hearst et al.~\cite{hearst2019toward} explored appropriate visualization responses to singular and plural superlatives (e.g., ``highest price'' and ``highest prices'') and numerical graded adjectives (e.g., ``higher'') based on data distributions. 

Gaba et al.~\cite{gaba2022comparison} took a different approach.
Having identified natural language associated with specific comparison types, the authors generated a mapping between natural language descriptions of comparison intends and best visualization design to facilitate those comparisons. 
Researchers also built systems and comparative visualization tools to support more specific comparison needs \cite{Gleicher2011}, such as systems to compare social networks \cite{pister2023combinet}, tools to support data comparisons in virtual realty \cite{joos2022visual}, and visualizations to reveal similarities and differences between two streams or sequences of multi-item data \cite{chen2020co}.

At the same time, empirical findings with human participants in these efforts to support comparison processes of visualized data also led to advances in the understanding of the perceptual and cognitive mechanisms for visual comparisons. 
For example, certain visual characteristics are easy to compare, such as object translations \cite{franconeri2021three} and data density \cite{pandey2016towards}.
Other visual encoding channels may complicate comparison tasks \cite{franconeri2013nature,larsen:1998, cleveland1984graphical}, such as orientation, texture, color, and scaling. 
These visual characteristics can lead to varying comparison affordances.
For example, people compare different pairs of data values depending on how they are spatially arranged (e.g., vertical alignment vs. horizontal alignment) \cite{lyi2020comparative, jardine2019perceptual, ondov2018face}, driving different takeaways \cite{xiong2021visual}.
Choices for visualization type can also influence what people compare and takeaway from data.
For example, people prefer comparing pair-wise data magnitudes when they read bar charts (e.g., this bar is taller than that bar), but for line charts, they prefer comparing changes over time or relations (e.g., x increases with time) \cite{xiong2021vss, zacks1999bars, xiong2019illusion, lee2016vlat, saket2018task}.
The number of data values presented can also influence the speed and accuracy of comparisons \cite{nothelfer2019measures, mccoleman2021rethinking}.
The more items to compare, the less efficient the comparison becomes. 

\subsection{Perception and Grouping in Visualizations}
To make data comparisons, viewers must first extract data groups and then compare them to identify differences or similarities
\cite{shah2011bar, michal2017visual, michal2016visual}.
The data group extraction is usually done by dividing data points in a visualization into visual units~\cite{pinker1990theory}.
As our visual system tends to perceive ``the forest before the trees''~\cite{navon1977forest}, grouping data elements is more intuitive than perceiving the individual data items ~\cite{ahissar2004reverse,love1999structural}.
Existing work in visualizations has generated models that predict which data elements people tend to group~\cite{fygenson2023arrangement, bearfield2023does}.
For example, when people read bar charts depicting two groups of two bars, they tend to group spatially proximate bars and compare the average height of those groups~\cite{bearfield2023does}. 

Perceptual features of the data points can influence how people form visual groups. 
These features are generally categorized as either `spatial' or `featural'~\cite{kim1999grouping,yu2014grouping, yu2019similarity}.
Featural cues can include color, shape, or size encoding.
 
Some features are more efficiently and intuitively grouped than others. 
For example, color similarity is a stronger grouping cue than shape similarity, such that clusters in a scatterplot are more accurately distinguished when they have distinct colors compared to distinct shapes~\cite{duncan1989visual, gleicher2013perception}. 
Within shape encodings, open shapes such as crosses and asterisks are harder to differentiate amongst other open shapes, compared to closed shapes such as squares and circles~ \cite{burlinson2017open}.
Spatial cues can include proximity (objects that are close to each other),  co-linearity (objects that are placed along a common line), and region-sharing (objects that are placed in the same region). 
Generally, the distance between data elements matters.  
Increased spacing between units affords separation of those two units~\cite{shah1999graphs, brooks2015traditional}.
Spacing out groups of bars with irregular distances in bar charts makes messages concerning those groups more salient than bar charts with uniform spacing~\cite{fygenson2023arrangement}.

Existing work has demonstrated that spatial cues often outweigh featural cues in influencing perception~\cite{franconeri2009number, yu2014grouping,yu2019gestalt, zhao2019neighborhood}.
Spatial cues are harder to inhibit, even when the viewer knows that grouping data units spatially can harm task performance \cite{brooks2015traditional, wagemans2012century}. 
Data visualizations tend to leverage both spatial and featural cues to encode data, but spatial grouping cues seem to get prioritized in visualization perception~\cite{bearfield2023does}.
Once the viewer selects one element to focus attention, they become more sensitive to nearby elements~\cite{posner1980,folk2010}.
They therefore more likely select more spatially proximate elements to compare, disregarding whether they share featural similarities or not~\cite{wagemans2012century}.
For example, a viewer will prioritize comparing two bars that are close to each other even if they have different colors, rather than prioritizing comparing bars that have the same colors but are further away from each other. 

Grouping cues significantly influence which elements users compare in a visualization.
If the grouping cues mislead the viewer to form the `wrong' visual groups, they will end up making invalid or ineffective comparisons. 
However, it is not easy to ensure that the viewers select the `right' groups because grouping perceptual features can compete against each other. 
The visualization community does not yet have comprehensive knowledge of how design features drive grouping behaviors in visual data interpretation \cite{fygenson2023arrangement}.
This can be dangerous because once visual groups are formed, it is extremely difficult to break those groups to form new groups (this process is referred to as perceptual reorganization)~\cite{bearfield2023does}.
Analysis of visual intelligence tests has demonstrated that perceptual reorganization tasks are amongst the most difficult problems on the test because they require very effortful perceptual computation~\cite{lovett2017modeling}. 
This suggests that chart viewers would encounter difficulty with perceptual reorganization.
Therefore it is critical to design visualizations to immediately and strongly support the formation of the `right' groups.
This motivate us to develop a model that can predict and constrain the visual groups chart viewers form pre-attentively. 

\section{Data-Induced Groupings}

We conducted two studies on Prolific~\cite{palan2018prolific} to understand how participants perceptually group data, and when these groupings may lead to misinterpretations.
Both studies used dot plots with a nominal x-axis and numeric y-axis.  The first asked participants to identify groups, and the second clarified the nominal nature of the x-axis.   We hypothesized that explicit instructions would reduce reliance on ordering or grouping along the x-axis.

We then trained interpretable machine learning models to predict whether a participant would identify a given subset of points in a chart as a group.  Using these models, we evaluated our hypotheses and proposed a visualization design tool that accounts for data-induced groupings.  

\subsection{User Studies}
\subsubsection{General Methods}

\begin{figure}
    \centering
    \includegraphics[width=\columnwidth, trim={0cm 0.5cm 0cm 0cm}]
    {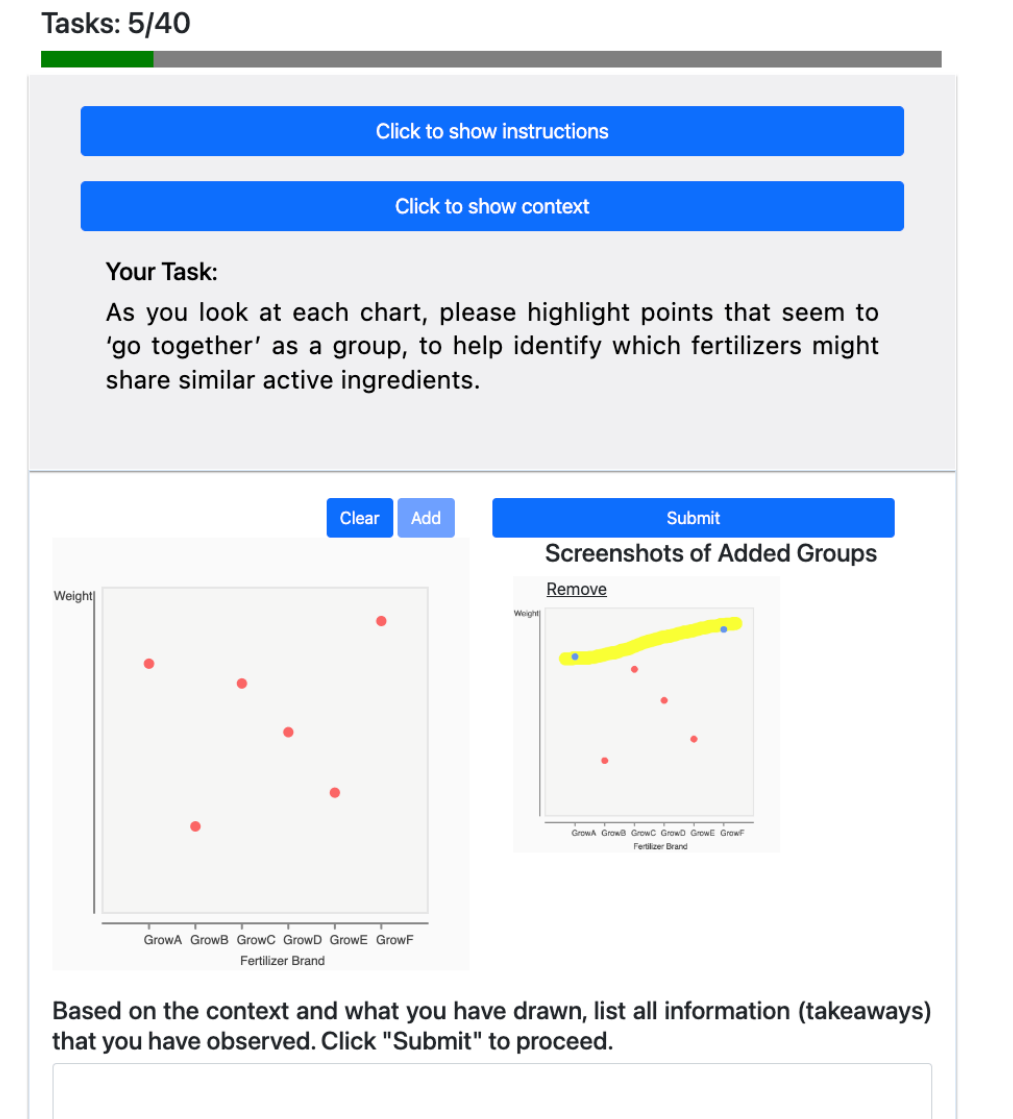}
    \caption{Task interface.  Participants can highlight or lasso groups of points, clear their selections, and add to or remove groups from the gallery on the right. After every 5 charts, we also add a text box so participants can describe what they see in the chart.}
    \label{fig:interface}
\end{figure}

Participants viewed a series of dot plots, each containing six points evenly placed on the x-axis, with randomly generated values between 0 and 100 for the y-axis, using the interface shown in \Cref{fig:interface}, for both studies. The major reason for choosing only six points is to strike a balance between cognitive load and participant engagement \cite{treisman1982perceptual}. 
Each study involved 40 randomized dot plots with y-values scaled for a fixed aspect ratio (5–95)~\cite{heer2006multi,wang2017there}.
\vspace{2mm}

\noindent\textbf{Study 1 Constellation (N = 20):} Users were asked to 
\textit{``highlight all groups of points that seem to `go together' on a given chart.''}
Overall, participants identified 2380 groups from 40 charts. We found that co-linearity was a strong predictor of group formation, as participants perceived trends as groups. Clustering was also influenced by separations on both x and y-axes, even though the x-axis was nominal and irrelevant. We named this the {\it Constellation} study due its similarity of using perceptual grouping principles to identify groups of constellations in the night sky~\cite{yantis1992multielement}.

\vspace{2mm}

\noindent\textbf{Study 2 Contextual (N = 20):} 
This study provided real-world context by associating the six points with performance scores (y-axis) for six fertilizer brands (x-axis). This framing was intended to make participants aware of the data context, encouraging them to form groupings based on perceived relationships in the data rather than solely on image-based patterns.
The instructions emphasized the nominal nature of the x-axis by referencing the fertilizers.
Participants were asked to hypothesize which data points might share a common hidden variable that associates them.  We also did not want to encourage or discourage any particular types of groupings.  Thus, we instructed as follows: \textit{``As you look at each chart, please highlight points that seem to `go together' as a group, to help identify which fertilizers might share similar active ingredients''}.    
Participants identified 2426 groups from 40 charts.
They were asked to write observations for every 5$^{th}$ chart, so we can gain a deeper insight into how participants formulated their groups. 
We called this the {\it Contextual} study.

\vspace{2mm}

Participants used the interface (\Cref{fig:interface}) to highlight or lasso the points that they consider a group. 
They can clear their selection, add a group to a gallery on the right side, and remove a group from the gallery. 
When they are satisfied with the groups they have added, they can click submit.   
 
To familiarize participants with the interface, a mandatory tutorial taught them how to highlight, delete, and clear a set of points on an example task; the tutorial used both text instruction and visual animations. 
Each participant was required to replicate the instructions on the example chart in order to proceed. 
At the end of the study, participants were asked a brief demographic survey (age group and education level). Participants were all compensated at a rate of \$12/hr.

To minimize malicious behavior on the crowdsourcing platform~\cite{borgo2017crowdsourcing, borgo2018information}, we implemented several measures. We pre-screened participants on Prolific for English fluency and a minimum 97\% past approval rate. Task completion times were recorded, and outliers were excluded. 
Finally, we conducted a series of pilot studies prior to launching the main experiment. Additionally, 21 participants took part in pilot studies using the same stimuli and interface as the main study. The pilot results informed the design of tutorial sections to ensure participants could follow the experiment effectively.

\yilan{Out of 61 recruited participants (20 in Study 1, 20 in Study 2, and 21 in a pilot), 47 provided complete demographic questionnaires. Their mean age was 26.4 years (SD = 5.9, range 19-41). Just under half held at least a bachelor's degree (21/47 $\approx 45$\%), whereas the remainder reported an associate degree, some college, or a high-school diploma (55\%). Prior data-visualization experience, rated on a fix-point Likert scale (1 = none, 5 = frequent creator), averaged $2.77 \pm 1.34$, indicating low-to-moderate familiarity. Self-reported effort during the task, on a three-point scale (1 = very little, 3 = high), averaged $2.45 \pm 0.62$, suggesting that participants generally engage in the tutorial and grouping tasks. }

\subsubsection{Hypotheses}

We made two hypotheses for the study.
While similar, the first seeks to verify that participants rely less on trend shapes when the x-axis is known to be nominal, while the second seeks to identify what participants do rely on:

\stitle{H1:}  In the Contextual study, participants will rely less on co-linearity compared to Constellation study. 

\stitle{H2:}  In the Contextual study, participants will focus more on y-axis values compared to the Constellation study. 

\subsection{Overview of User Study Results}
\label{ss:userstudyoverview}
Before we dive into model development, we first present an overview of the user-selected groups.   We organize these into basic statistics about the groups, and then semantic features about the groups.

\begin{figure}
    \centering
    \includegraphics[width=\columnwidth, trim={0cm 0.5cm 0cm 0cm}]{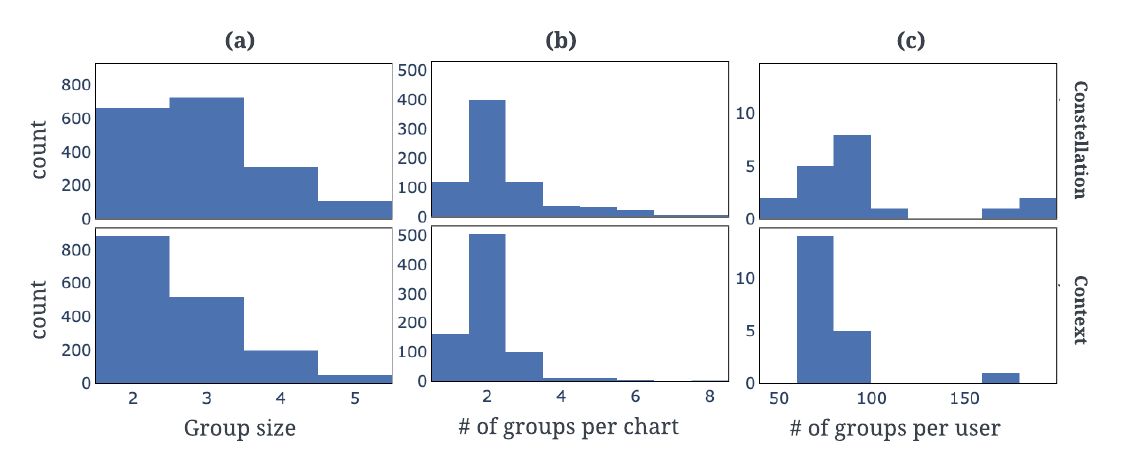}
    \caption{(a) Distribution of group sizes, (b) distribution of the number of groups per chart, and (c) distribution of the number of groups per participant.}
    \label{fig:group_distributions}
\end{figure}

\subsubsection{Basic Statistics}
We first report the number and sizes of groups overall, and on a per chart/participant basis, for both studies in \cref{fig:group_distributions}.   The Constellation study exhibited larger group sizes, more groups per chart, and more groups per participant than the Contextual study. In both studies, and particularly in the Contextual study, the number of groups per chart is concentrated around two. This makes sense because participants likely relied only on the y-axis for form groups, given that there were only six points in total, with most groups consisting of two to three points.

\subsubsection{Semantic Statistics}

To study the nature of the groups, we manually examined a large sample of groups and designed two classes of features based on common  properties we observed: co-linearity and clustering.
 The co-linearity features are computed by fitting a linear model to the input group, and calculating its error (sum of residuals) and slope.  
The clustering features follow the intuition that participants identify clusters of points based on their distance from the other points in the chart. 
Forming clusters is an intuitive perceptual process that commonly occurs in visual analysis, especially with dot plots~\cite{amar2004,yantis1992multielement,julesz1969cluster}.
Points that are spatially proximate or share similar visual features (e.g., the same color), tend to be grouped together. 
\begin{figure}
    \centering
    \includegraphics[width=\columnwidth, trim={0cm 0.5cm 0cm 0cm}]{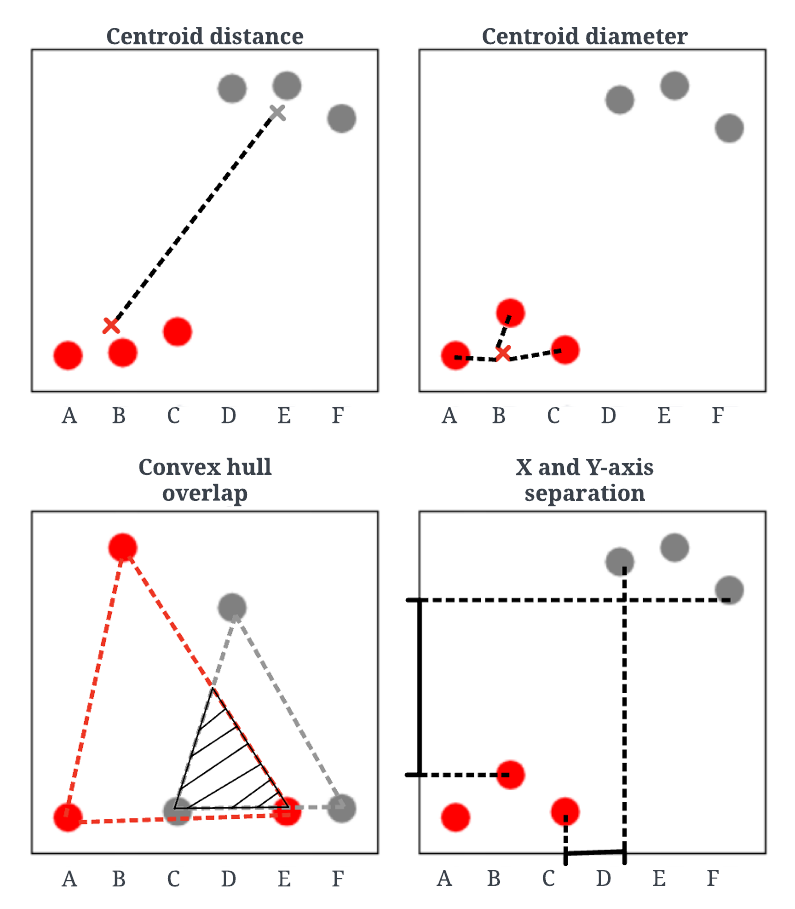}
    \caption{Examples of clustering features except for cluster ratio, which is the ratio between centroid distance and cluster diameter.  The red colored dots denote the candidate group $g$. }
    \label{fig:features}
\end{figure}
%
One challenge is that there are many ways to evaluate cluster separation between a target group of points $g$ and the rest of the points in the chart $r$. We ultimately designed several common features, described below and summarized in \Cref{fig:features}:

\begin{itemize}[leftmargin=*, itemsep=0in]
    \item \stitle{Centroid distance}: We compute the Euclidean distance between the centroids of $g$ and $r$, and normalize by dividing by the diagonal length of the chart.
    \item \stitle{Centroid diameter}: We compute the average Euclidean distance between each point in $g$ and its centroid. 
    \item \stitle{Centroid ratio}: We take the ratio of centroid distance to centroid diameter.
    \item \stitle{Convex hull overlap}: Given the convex hull $h_g$ of $g$ (and $h_r$ of $r$), we compute the ratio between their perimeter of overlap and union of their areas.  For convex hulls that form a line, we assume that the line has a width of 1 pixel:   
    $(h_g\cap h_r)/(h_g\cup h_r)$.
    \item \stitle{X- and Y-axis separation}: When the x-axis is nominal, separation along the x-axis should {\it not} be a factor for identifying groups.  Therefore, we compute a separation measure for each axis in isolation.  For the y-axis, feature $f_y$ measures the minimum distance between points in $g$ and points $r$.  Similarly, $f_x$ is defined along the x-axis.
    We expect $f_y$ to be the dominant feature; when the x-axis is nominal, co-linearity and $f_x$ should not influence visual groupings.  
    
\end{itemize}

\begin{figure}
    \centering
    \includegraphics[width=\columnwidth, trim={0cm 0.5cm 0cm 0cm}]{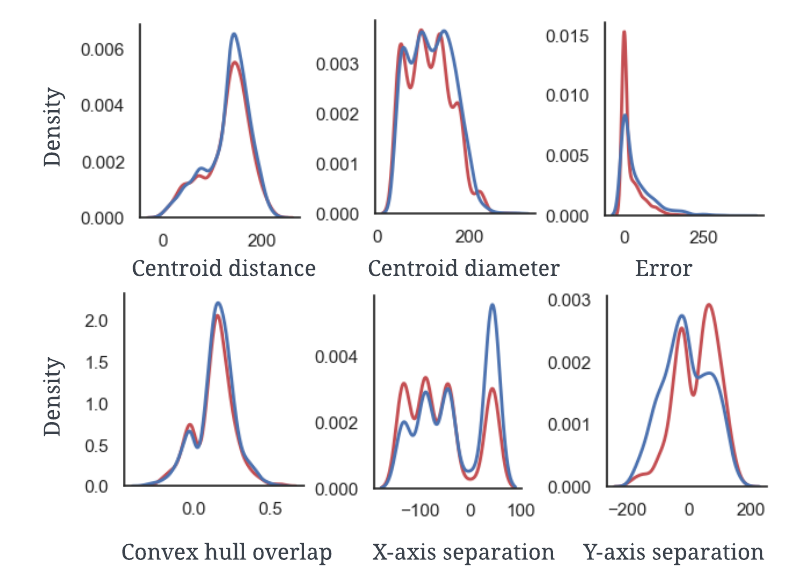}
    \caption{Distributions of centroid distance, centroid diameter, error, convex hull overlap, x and y-axis separations. The \blue{Constellation study is blue} and the \red{Contextual study is red}.}
    \label{fig:feature_distributions}
\end{figure}


\begin{table}[t]
    \centering
    \renewcommand{\arraystretch}{1.2}
    \caption{Fixed-effect estimates for the \texttt{study} factor (Constellation vs.\ Contextual).}
    \label{tab:mixed_effects_results}
    \begin{tabular}{l rrrr}
        \toprule
        \textbf{Dependent Variable} & $\boldsymbol{\beta_1}$ & \textbf{SE} & \textbf{$z$} & \textbf{\textit{p}} \\
        \midrule
        \textbf{Error}              &  15.630 & 5.523 &  2.830 & $\mathbf{0.005}$ \\
        Centroid diameter           &   2.407 & 6.649 &  0.362 & 0.717 \\
        Centroid distance           &   2.821 & 3.959 &  0.713 & 0.476 \\
        Convex-hull overlap         &   0.013 & 0.012 &  1.085 & 0.278 \\
        \textbf{X-axis separation}  &  29.578 & 5.990 &  4.938 & $\mathbf{<0.001}$ \\
        \textbf{Y-axis separation}  & $-33.049$ & 8.481 & $-3.897$ & $\mathbf{<0.001}$ \\
        \bottomrule
    \end{tabular}
\end{table}

The overall density distributions of each feature are depicted in \cref{fig:feature_distributions}. Features such as centroid distance, centroid diameter, and convex hull overlap exhibit similar distributions in both studies. Notably, in the Contextual study, errors are more concentrated around 0, aligning with the higher prevalence of 2-point groups, as illustrated in \cref{fig:group_distributions}. The Constellation study also tends to feature more groups positioned farther away from other points along the x-axis, whereas the Contextual study tends to exhibit more groups positioned further away along the y-axis.

\yilan{To assess whether the six semantic metrics differ by study condition, we fit separate linear mixed‐effects models with study (\emph{Constellation} vs.\ \emph{Contextual}) as the fixed effect and a random intercept for participant (40 charts per participant). Table~\ref{tab:mixed_effects_results} reports the fixed‐effect coefficients~(\( \beta_1 \)), which quantify the mean shift relative to the Contextual baseline. Constellation charts exhibit significantly \emph{higher error} and larger \emph{x‐ and y‐axis separations} (all \( p<.01 \)); centroid diameter, centroid distance, and convex‐hull overlap show no reliable differences. These inferential results reinforce the descriptive trends in Fig.~\ref{fig:feature_distributions}.}. 

\subsection{Model Development}

To analyze the results of each study, we designed a predictive model that takes as input a set of points and outputs the likelihood that a participant will select them as a group. One challenge is that the number of points in a group may vary, so they cannot be directly used as model inputs.  
In addition, we aimed to develop an interpretable model with simple features. 

\stitle{Models and Features.} For these reasons, we used the semantic features from \Cref{ss:userstudyoverview} 
and evaluated them using logistic regression and decision tree models.  
We tested convolutional neural networks (CNNs), but found that their performance was comparable to simpler decision trees (F1: 96–97\%). Given the small dataset and similar accuracy, we prioritized lightweight, transparent models.

\stitle{Training Data.}
One challenge with fitting a model is that the study data can suffer from class imbalance because participants do not always select the non-groups.  We cannot trivially label all subsets of non-selected points as negatives.
For example, a participant might select points AB to be a group, and might consider CDE as another potential group but choose not to label it. 

Therefore, we sought to identify subsets of points in each chart that were clearly not groups.  Specifically, let $g$ be a candidate subset of points, we then assessed it based on three criteria: 1) $g$ was not selected by any participant, 2) its linear fit was poor relative to the groups selected by participants, and 3) $g$'s convex hull did not overlap with the convex hull formed by the rest of the points in the chart.  The second criteria was measured using the linear fit's error (sum of residuals), and we checked that $g$'s error is larger than the average error.   The third criterion simply checked that the convex hull feature was zero.

Based on these requirements, we generated 1272 examples to validate the negative examples. 
One of the authors manually labeled these examples to identify ``false'' negative ones.
For this process, as shown in Figure~\ref{fig:additional_feature}, we categorized all the instances where red points \textit{did} form a group as a ``false'' negative example. 
We only identified 42 false negative examples out of the 1271 instances, indicating that our negative example generation procedure was valid. 

Since each chart contained only six points and a group of one does not make sense, we limited our training set to groups of 2-5 points.   
In summary, the Constellation study contains 1502 (1308) positive (negative) examples, and the Contextual study contains 1639 (1396) positive (negative) examples. For each study, we used 70\% of the data for training and hyper-parameter tuning, 20\% for testing, and the remaining 10\% as holdout data for reporting final results.

\subsection{Model Training and Evaluation}

\begin{figure}
    \centering
    \includegraphics[width=0.95\linewidth, trim={0cm 0.5cm 0cm 0cm}]{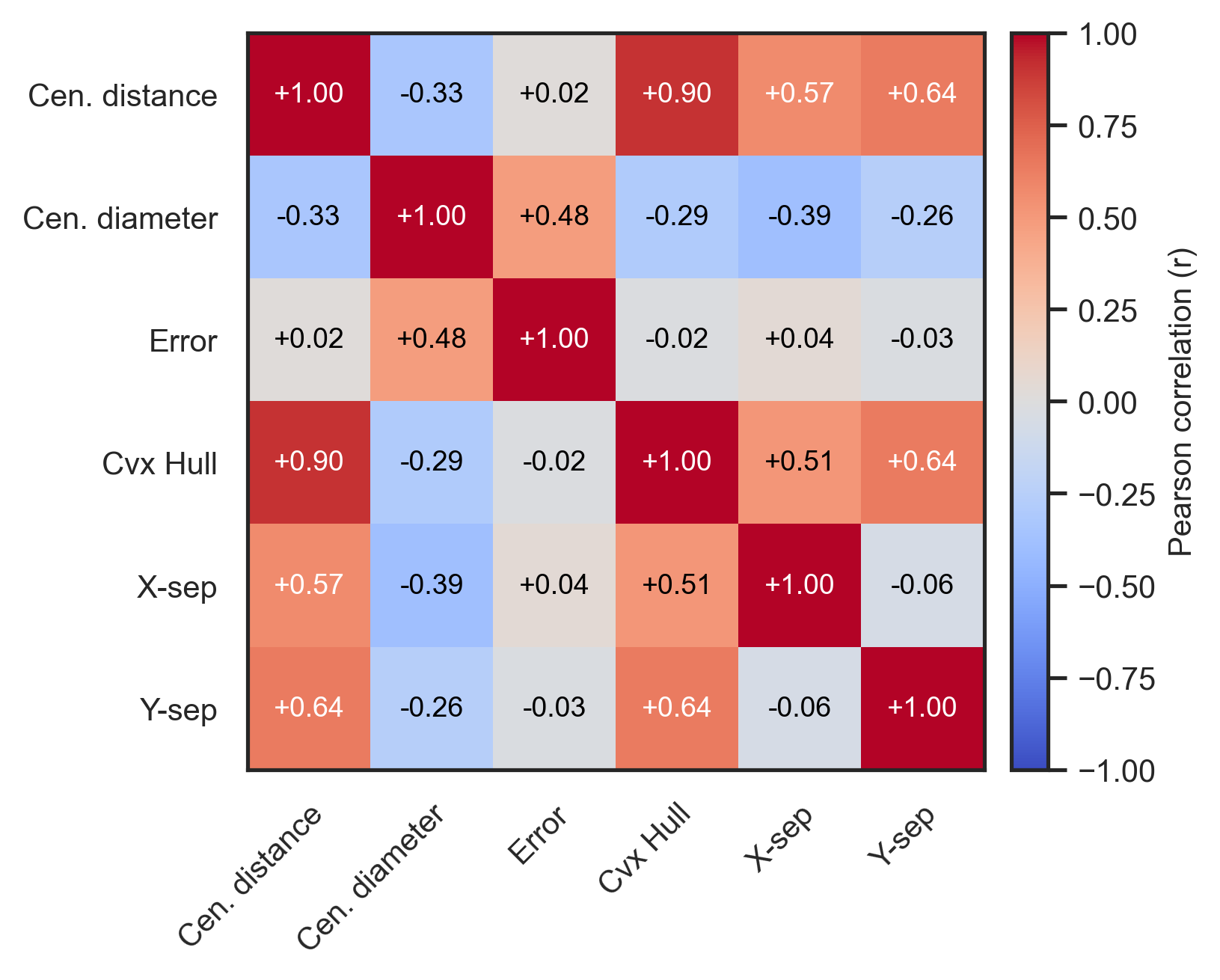}
    \caption{\yilan{Pair-wise Pearson correlation matrix for the six semantic features in the Constellation study. Cell colours indicate correlation strength and sign (red = positive, blue = negative); the numeric labels give $r$ rounded to two decimals.
Abbreviations—Cen. distance: centroid distance; Cen. diameter: centroid diameter; Error: grouping error; Cvx Hull: convex-hull overlap; X-sep: x-axis separation; Y-sep: y-axis separation.}}
    \label{fig:corr}
\end{figure}

We trained separate logistic regression and decision tree models, \yilan{with the decision trees limited to a maximum depth of 3, using data from each user study and all available features.} For logistic regression, we used Variance Inflation Factor (VIF) to eliminate highly correlated features. Notably, centroid diameter and centroid distance exhibited a high correlation with convex hull overlap and x-separation, so we removed the former two. \yilan{Figure \ref{fig:corr} presents the Pearson correlation matrix for the six semantic features in the Constellation study, highlighting their pairwise interdependencies.}

\yilan{To improve readability, we use abbreviated metric names in the tables: \textbf{P} for \textit{Precision} and \textbf{R} for \textit{Recall}. These abbreviations are used consistently throughout the paper.} 
\yilan{\Cref{tab:evaluation1} reports the mean and standard deviation of the five-fold cross-validation (CV) testing scores together with the results on the independent hold-out (HO) set. 
On the HO data,} \Cref{tab:evaluation1} shows that all models have 100\% precision, meaning any groups it identified were identified as groups by participants.  This suggests reliability in its positive predictions.   The decision tree model demonstrated improved recall of $5-9\%$ compared to logistic regression, likely due to non-linear relationships between the features and target variable. Finally, the decision tree's F1 scores ranged from $96\%$ to $97\%$ across both studies, highlighting how even simple decision tree models can effectively identify data-induced groupings. 



\begin{table*}[t]
    \centering
    \small                                        
    \setlength{\tabcolsep}{4pt}                   
    \renewcommand{\arraystretch}{1.15}
    \caption{Five-fold cross-validation (CV) and independent hold-out (HO) performance.  
             Each entry reads “CV mean ± SD / HO”.  Bold F1 marks the best model per study.}
    \label{tab:evaluation1}
    \begin{tabular}{l rrr rrr}
        \toprule
        & \multicolumn{3}{c}{\textbf{Logistic Regression}} 
        & \multicolumn{3}{c}{\textbf{Decision Tree}} \\
        \cmidrule(lr){2-4}\cmidrule(lr){5-7}
        & \textbf{P} & \textbf{R} & \textbf{F1}
        & \textbf{P} & \textbf{R} & \textbf{F1} \\
        \midrule
        Constellation 
          & 0.94 $\!\pm\!$ 0.02 / 1.00 
          & 0.88 $\!\pm\!$ 0.02 / 0.87 
          & 0.91 $\!\pm\!$ 0.01 / 0.93 
          & 0.97 $\!\pm\!$ 0.03 / 1.00 
          & 0.88 $\!\pm\!$ 0.03 / 0.95 
          & \textbf{0.92 $\!\pm\!$ 0.03 / 0.97} \\
        Context       
          & 0.94 $\!\pm\!$ 0.01 / 1.00 
          & 0.91 $\!\pm\!$ 0.02 / 0.86 
          & 0.92 $\!\pm\!$ 0.01 / 0.92 
          & 0.92 $\!\pm\!$ 0.01 / 1.00 
          & 0.94 $\!\pm\!$ 0.01 / 0.91 
          & \textbf{0.93 $\!\pm\!$ $\mathbf{<0.01}$ / 0.95} \\
        \bottomrule
    \end{tabular}
\end{table*}

Then we used SHapley Additive exPlanations (SHAP)~\cite{lundberg2017unified} to estimate the contribution of each feature to the final outcome. As shown in Eq.\ref{eq_shap} where $F$ is the set of all features, SHAP computes the marginal contribution of a feature $i$, denoted by $\phi_i$, by averaging its weighted contribution across all possible combinations. 
Because SHAP is model-agnostic, SHAP values serve as indicators of ``feature importance,'' enabling comparison of feature importance variations between logistic regression and decision tree models. 

\begin{equation}
\centering
\phi_i = \sum_{S\subseteq F\setminus \{i\}}\frac{|S|!(|F|-|S|-1)!}{|F|!}[f_{S\cup\{i\}}(x_{S\cup\{i\}})-f_{S}(x_S)]
\label{eq_shap}
\end{equation}

Table \ref{tab:models} presents the absolute feature importance values in both models and studies. For instance, the SHAP value of slope implies that the marginal contribution of a higher slope to the final outcome of the model is $0-0.01$ on average. 
The SHAP value can also be computed on a per-sample basis; however, we reported the average across samples for this study.

\begin{table*}[t]
    \centering
    \caption{Feature importance (mean absolute SHAP values) for logistic-regression (LR) and decision-tree (DT) models.  
             Abbreviations: X-sep = X-separation, Y-sep = Y-separation, Cvx Hull = Convex-hull overlap, Cen.\ Ratio = Centroid ratio.}
    \label{tab:models}
    \begin{tabular}{llrrrrrr}
        \toprule
        & & \multicolumn{6}{c}{\textbf{Feature Importance (|SHAP|)}} \\
        \cmidrule(lr){3-8}
        & & Slope & Error & X-sep & Y-sep & Cvx Hull & Cen.\ Ratio \\
        \midrule
        Constellation & LR & 0.01 & 0.00 & 0.06 & 0.03 & \textbf{0.31} & \textbf{0.22} \\
        Constellation & DT & 0.00 & \textbf{0.19} & 0.00 & 0.00 & \textbf{0.15} & \textbf{0.23} \\  
        Context       & LR & 0.01 & \textbf{0.09} & 0.02 & \textbf{0.10} & \textbf{0.14} & \textbf{0.18} \\
        Context       & DT & 0.00 & \textbf{0.15} & 0.00 & \textbf{0.13} & \textbf{0.16} & \textbf{0.11} \\
        \bottomrule
    \end{tabular}
\end{table*}

    

\begin{figure}
   \centering
   \includegraphics[width=\columnwidth, trim={0cm 0.5cm 0cm 0cm}]{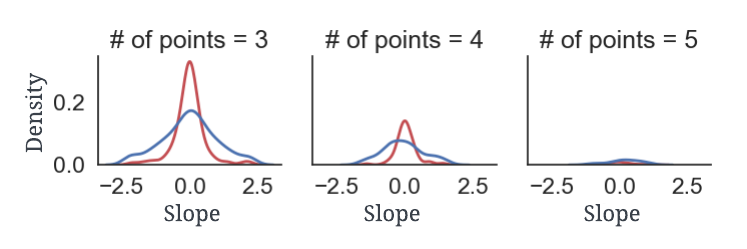}
   \caption{Distribution of slope when the the linear fit is tight (error is low) by group size. The \blue{Constellation study is blue} and the \red{Contextual study is red}.  Lower error means tighter linear fit.  When a group has a tight linear fit (lower error, left chart),  the Contextual study's distribution is narrowly centered around zero.  This means they have near-horizontal slope as compared to the Constellation study.    }
   \vspace{-.2in}
   \label{fig:slopes-overview}
\end{figure}

\subsubsection{Analysis}

We found that convex hull overlap, centroid ratio, and co-linearity (as measured by linear fit {\it Error}) had the strongest influence on group identification.  In combination with the high model accuracy, their significant contribution supports our intuition that cluster separation and co-linearity play large roles in predicting data-induced groupings.   

The slope of the linear line did not affect group identification. In contrast, linear fit error showed high contribution even in the Contextual study where the instructions emphasize that the x-axis is nominal. This suggests that data-induced groupings can overpower the background knowledge users may have about the chart.  Since slope did not have any effect in our experiments, we used the term {\it co-linearity} and its error feature synonymously in the rest of this paper.

When comparing single-dimension features, we found that users did not rely solely on x-axis separation to identify groups in either study.  At the same time, y-separation showed near-zero contribution under the Constellation study, but its contribution increased to $0.1-0.13$ in the Contextual study.  This is nearly as high as the contributions of co-linearity and clustering features, suggesting that users attempt to ignore the effects of the x-axis and instead focus on separation along the y-axis.

\begin{figure}
    \centering
    \includegraphics[width=0.75\columnwidth, trim={0cm 0.5cm 0cm 0cm}]{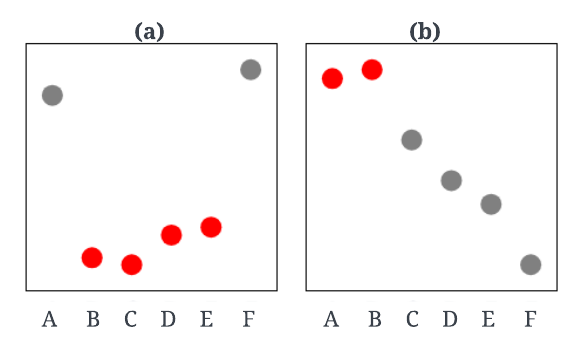}
    \caption{Low slope examples. The \red{chosen group $g$ is red} and the \gray{rest of the points $r$ are grey}. $g$ is nearly horizontal and far away from $r$, thus it could be interpreted as a cluster rather than line.  }
    \label{fig:lowslopexmple}
\end{figure}

\subsubsection{Evaluating Hypothesis 1}
Hypothesis 1 asks whether users rely less on co-linearity under the Contextual condition as compared to the Constellation condition.   
At first glance (\cref{tab:models}), the SHAP results appear to suggest that this is not the case: 
co-linearity's contribution is higher under the Contextual condition than the Constellation one, indicating that users seem to rely {\it more} on co-linearity in the Contextual condition.   

We observed that under the logistic regression for the Constellation condition in \cref{tab:models}, co-linearity surprisingly had zero contribution, while convex hull's contribution was roughly the sum of error and convex hull in the Contextual condition.  This disparity is unlikely to be natural since co-linearity generally has a high contribution when switching from a linear to decision tree model class.   

\yilan{This suggests that co-linearity and convex-hull overlap encode partially redundant information. Because of sampling variability, the decision tree happened to rely more on convex-hull overlap, which in turn skewed the SHAP attributions.}
We investigated this by training  two decision tree models for each study: one using only co-linearity features and the other using only clustering features.   We found that 77\% (81\%) of the test samples in the Constellation (Contextual) study were accurately predicted using both models.   The large overlap suggests that when using both sets of features for training, the model may select either.

We further plotted the distribution of slopes for all groups with low linear fit error, faceted by the size of the group (\Cref{fig:slopes-overview}).  We excluded 2-point groups since every pair of points trivially forms a line.   
%
A  distinctive pattern immediately emerged: the distribution of near-horizontal slopes in the Contextual condition showed considerably higher density than in the Constellation condition.   These groups largely comprised set that were well-predicted by both the co-linearity-only and cluster-only models. This pattern is evident in  \Cref{fig:lowslopexmple}, which plots two example groups with low slope.  Both groups could reasonably be categorized as clusters rather than trends as they were separated from the rest of the points along the y-axis (high y-separation).    

\begin{figure}
    \centering
    \includegraphics[width=0.75\columnwidth, trim={0cm 0.5cm 0cm 0cm}]{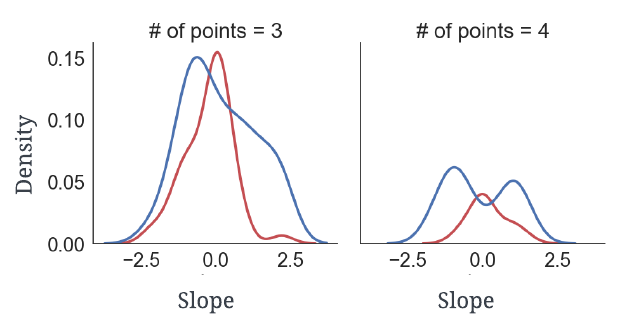}
    \caption{Distribution of slope for groups which are mispredicted by the cluster-only model. The \blue{Constellation study is blue} and the \red{Contextual study is red}.  Lower error means tighter linear fit. Note, there are no slope distribution for group size of 5 because none of these groups are mispredicted by the cluster-only model.  }
    \label{fig:slopedensity_after}
\end{figure}


To validate this hypothesis, we used a model cascade where training samples were first fed to a decision tree model that used only clustering features.   Only the incorrectly predicted  samples were fed into a second decision tree model trained on co-linearity features.   This approach enables us to isolate the effect of co-linearity from clustering.

\Cref{fig:slopedensity_after} re-plots the distribution of slopes for only the groups mispredicted by the cluster-only model, faceted by group size.  All groups of size 5 were correctly predicted by the cluster-only model, so we did not plot their distribution.    We observed that the peak in the Contextual condition disappeared and became comparable to or lower than the  distribution under the Constellation condition.    This indicates that the majority of co-linear groups with near-horizontal slope were identified as clusters.   

\begin{figure}
    \centering
    \includegraphics[width=.75\columnwidth, trim={0cm 0.5cm 0cm 0cm}]{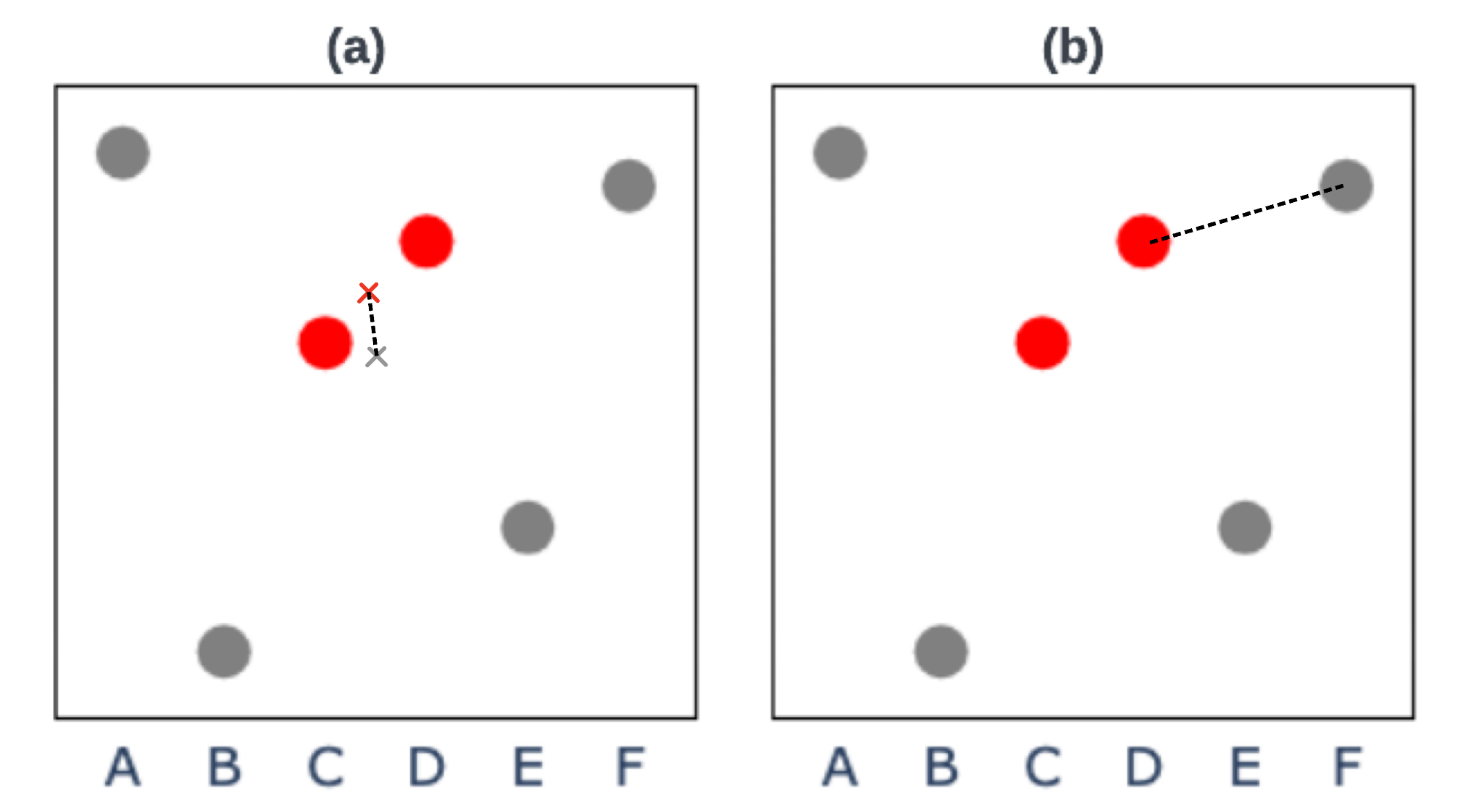}
    \caption{Example of false negative group from the Contextual study. The chosen group \red{$g$ is colored red}, and the \gray{rest of the points in the chart $r$ are colored grey}. (a) $g$ is located within the convex hull formed by $r$, and the distance between their centroids ($\times$ marks) is very small. (b) instead of computing the centroid ratio, we use the minimum euclidean distance between points in $g$ and $r$, depicted as a dashed line.}
    \label{fig:additional_feature}
\end{figure}
Furthermore, the Constellation study's distribution of slopes was much wider: a large subset of 3-point groups in the Constellation study had slopes of $0.25-2.5$, while the 4-point groups exhibited a bimodal distribution with most groups having slopes of $-1$ or $1$.  In contrast, the Contextual condition's distribution remained centered at 0 with a steep drop-off.   

Finally, \Cref{tab:evaluation3} reports the accuracy of the second model applied only to the test samples that the clustering features could not predict.  A higher F1 score indicates that co-linearity is more predictive for non-cluster groups. We found that the Constellation condition achieved a much higher F1 score ($0.91$) compared to the Contextual condition ($0.85$), suggesting that users rely less on co-linearity when a set of points is not an obvious cluster. It is important to note that using F1 scores for inference should be interpreted as indicative of general trends rather than as statistically validated evidence.

Based on this analysis, we believe there is weak evidence that users rely slightly less on co-linearity under the Contextual condition, although they still rely on it to a large degree.  


\begin{table}[t]
    \centering
    \renewcommand{\arraystretch}{1.2}
    \caption{Evaluation of the second stage in the model cascade.  
             The decision tree uses only co-linearity features and is applied to samples misclassified by the cluster-only model.}
    \label{tab:evaluation3}
    \begin{tabular}{l rrrr}
        \toprule
        & \textbf{P} & \textbf{R} & \textbf{F1} & \textbf{Support} \\
        \midrule
        Constellation & 0.91 & 0.92 & 0.91 & 151 \\
        Context       & 0.81 & 0.89 & 0.85 & 112 \\
        \bottomrule
    \end{tabular}
\end{table}

\begin{figure}
    \centering
    \includegraphics[width=\columnwidth, trim={0cm 0.5cm 0cm 0cm}]{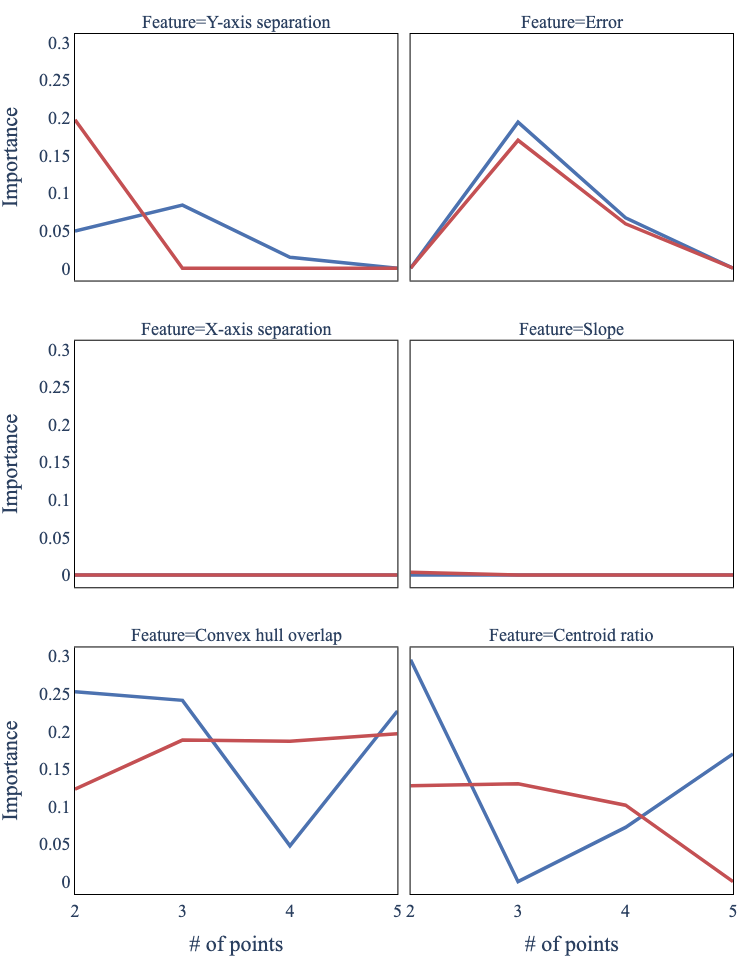}
    \caption{Group size vs SHAP feature importance on decision tree models. The \blue{Constellation study is in blue} while the \red{Contextual study is in red}. The y-axis reports average absolute SHAP values, which represent the marginal contribution of the target feature to the model's prediction. }
    \label{fig:feature_number_of_pts}
\end{figure}

\subsubsection{Evaluating Hypothesis 2}

Hypothesis 2 examines whether users tend to rely more on y-separation compared to Constellation-based analysis.  To assess this hypothesis,  we individually trained decision tree models that used only y-separation as the input feature. A higher F1 score suggests that y values alone are predictive of whether a set of points is considered a group. We found that switching from the Constellation to Context studies increased the F1 score on the hold-out dataset from $83\%$ to $92\%$. This supports our hypothesis. 

\subsection{Further Model Refinements}\label{ss:erroranalysis}

The previous study suggests that group size may play an important role in the model.  We now analyze the model's sensitivity to group sizes, and address a failure case related to the definition of the centroid ratio. Our results improve the model's F1 score to $0.97-0.99$ on the holdout.

\subsubsection{Group Size}

We first trained a decision tree model for each group size and plotted each feature's importance  as a function of group size (\Cref{fig:feature_number_of_pts}).  Interestingly,  features important for small groups (e.g., y-separation) are irrelevant for larger groups, while co-linearity is irrelevant for the smallest (2) and largest (5) group sizes, but becomes imporant for intermediate group sizes.
At the same time, cluster separation features are important at the smallest and largest group sizes but are
(relatively) less important for intermediate group sizes.   
Slope and x-separation are irrelevant across all cases. 

Overall, feature importance in the ``edge'' groups  (the smallest and largest group sizes) is consistently different than ``intermediate'' groups (3- and 4-point groups). For instance, co-linearity is not semantically meaningful for any two points that form a line, and a 5-point group contains all but one point in the chart, making its cluster separation from the remaining point more significant. Based on these findings, we trained two separate models: one for the edge groups and another for the intermediate groups. During inference, we selected the appropriate model based on the group size.

\subsubsection{Centroid Ratio}

To delve deeper into the factors contributing to lower recall rates for edge groups, we conducted an error analysis by visually inspecting all false negative samples within the testing datasets. Through this analysis, we uncovered a pattern that our current features fail to capture. Specifically, the model struggles to identify $g$ as a group if its points are entirely encompassed by the rest of the points $r$.  \Cref{fig:additional_feature} illustrates this case, where the points in $g$ are marked in red.

We found that this condition distorts the centroid ratio feature enough to cause misclassification.   
Specifically, the centroids of $g$ and $r$ are very close, and since the numerator for the centroid ratio is their centroid distance, the ratio ends up being near zero.

To address this challenge, we modified the centroid ratio by replacing the numerator with the minimum distance between points in $g$ and points in $r$.  This measure is consistent with centroid distance when the $g$ and $r$ are non-overlapping, and remains meaningful when $g$ is contained within $r$.    
We depict this in \Cref{fig:additional_feature}(b).  

\subsubsection{Putting It All Together}
The combination of these methods results in near-perfect models.  
\Cref{tab:evaluation2} summarizes the final model quality, where the overall F1 scores are $0.99$ and $0.97$ for the Constellation and Contextual studies, respectively.    For the Contextual study, the improvements primarily stem from segregating the edge and intermediate conditions.   By doing so, the edge (intermediate) models improved from $0.93$ to $0.95$ ($0.99$).  The original Constellation model already achieved a very high F1 score of $0.98$, so the edge (intermediate) models showed marginal improvements, increasing to $0.98$ ($0.99$).


\begin{table}[t]
    \centering
    \renewcommand{\arraystretch}{1.2}
    \caption{Final decision-tree (DT) performance on each study’s hold-out set.  
             Scores left of $\to$ are from the initial model; scores right of $\to$ are from the final model.  
             Precision is 1.00 in all cases because the hold-out sets contain no false positives.}
    \label{tab:evaluation2}
    \begin{tabular}{l rrr}
        \toprule
        & \textbf{P} & \textbf{R} & \textbf{F1} \\
        \midrule
        Constellation & \(1.00 \to 1.00\) & \(0.95 \to 0.97\) & \(0.97 \to 0.99\) \\
        Context       & \(1.00 \to 1.00\) & \(0.92 \to 0.94\) & \(0.96 \to 0.97\) \\
        \bottomrule
    \end{tabular}
\end{table}

\section{Potential Applications}\label{s:apps}

\yilan{In addition to theoretical applications, we illustrate how a visualization design might incorporate the diagnostic and redesign tools into their workflow, from identifying problematic groupings to evaluating alternative layouts. }Although data-induced groupings are often unintended artifacts when the x-axis is nominal (called hallucinations by Kindlemann et al.~\cite{kindlmann2014algebraic}), they are not inherently good or bad.    A designer who wishes to accentuate a set of desired groupings may wish to re-order the nominal axis so that those desired groups are more visually salient.  
Similarly, data-induced groupings that are not desirable may need to be ``broken'' by reordering the x-axis.  
Consider the following example:

\begin{figure}
    \centering
    \includegraphics[width=\columnwidth, trim={0cm 0.5cm 0cm 0cm}]{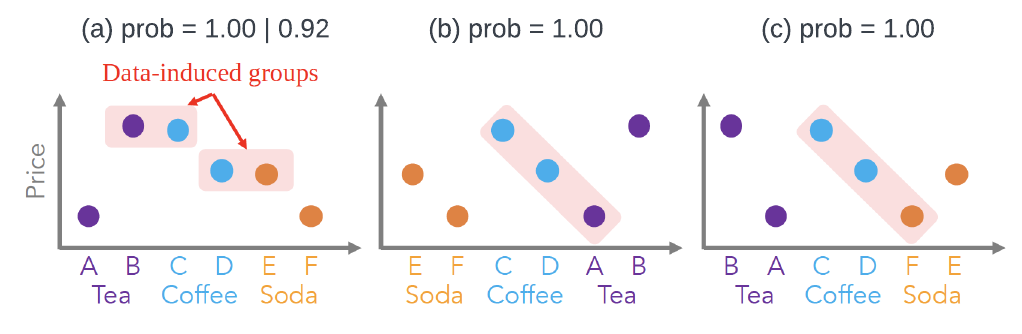}
    \caption{The  charts plot prices for items in different beverage categories.  Chart (a) exhibits data-induced groups of items that cross category boundaries, along with the trained model's predicted probability for those groupings.  Naively reordering the categories (b) or items within each category (c) can break those groupings but potentially induce new groupings (e.g. co-linearity).  }
    \label{fig:redesign}
\end{figure}

\begin{example}
\Cref{fig:redesign} plots prices for items in different beverage categories, where the beverage category forms a hierarchy that naturally groups the underlying data.  The designer wants to design a chart that minimizes comparisons across beverage hierarchies.   

\yilan{In realistic workflow, a designer creating a pricing dashboard could use the diagnostic tool to quickly check for misleading cross-category groupings before finalizing a chart for stakeholders.}
For instance, the left chart marks two data-induced groupings that span categories: BC groups tea and coffee items, while DE groups coffee and soda items.   

%
The designer might attempt to address these violations by swapping the Tea and Soda categories (middle) or by swapping B with A and F with E within their categories (right).   Although these adjustments break the above violations, they induce {\it new} cross-category groupings due to co-linearity.  
Being able to see these new groups would help the designer make the appropriate trade-offs.
Further, a redesign tool could automatically evaluate different permutations of the x-axis and recommend those that minimize violations and adhering to the hierarchy constraints.
\end{example}

This section shows how the predictive model can aid designers in achieving these goals.
We first describe a diagnostic tool that alerts the designer when a visualization is susceptible to data-induced groupings, such as those based on co-linearity.  
We then describe an automatic redesign tool that searches the space of nominal reorderings to emphasize desired groups and reduce the prevalence of undesired groups.

\begin{figure}
    \centering
    \includegraphics[width=\columnwidth, trim={0cm 0.5cm 0cm 0cm}]{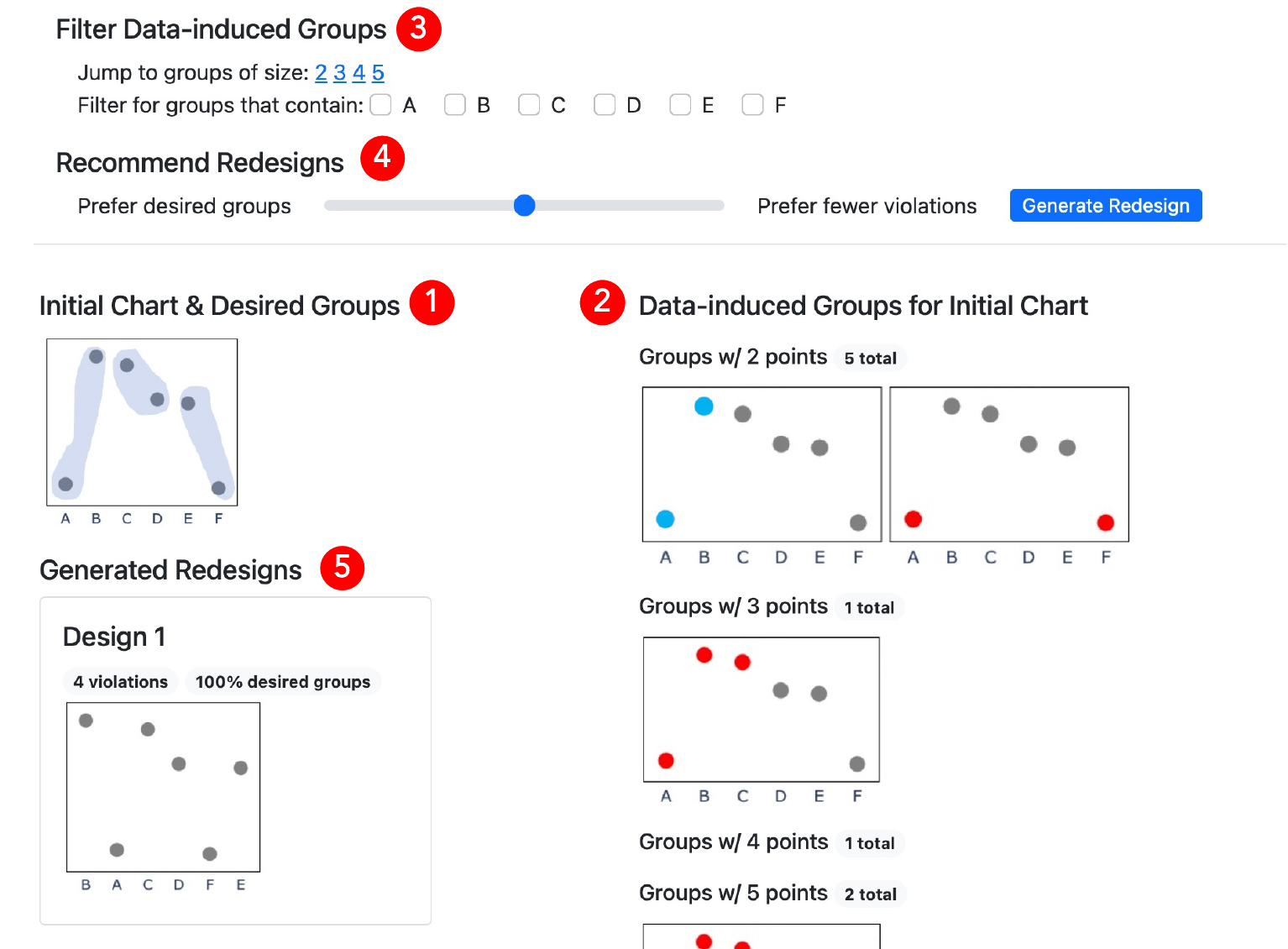}
    \caption{Interface for Diagnostic (1-3) and Redesign (4-5) tools.  The participant draws groups in an initial chart (1) and the system lists the data-induced groups (2), where desired groups are blue and violations are red.  Participants can filter the groups (3) or ask the system to propose a redesign (4,5). }
    \label{fig:app}
\end{figure}

\subsection{Diagnostic Application}

\Cref{fig:app} (1-3) presents the design of the diagnostic application.  
The main interface is split vertically, where the left side (1) shows the original chart (note that this is the same beverage dataset as in \Cref{fig:redesign}) along with the desired groups circled. 
The right sidebar (2) shows a gallery of the data-induced groups that the predictive model identifies.   Each group is shown as a separate chart, with the points in the group highlighted in red (undesired groups, also referred to as \textit{violations}) or blue (desired groups).   The charts are organized by group size, and the charts at each size are shown as a dense grid for easy scanning.   
The top of the interface (3) has controls to jump to a specific group size.
The user can also filter the groups to display those containing a specific set of points.

To identify data-induced groups, we developed a procedure $Diagnose(P, G_D)\to G_O$.  
It takes as input $P$, which is the set of $(x, y)$ pixel coordinates of the rendered points in the chart, and $G_D\subseteq P$, which is the set of desired groups. The procedure outputs the set of groups $G_O\subseteq P$ that the user is likely to perceive.  
Each element $g_o\in G_O$ is annotated with its uncertainty $g_o.prob$ as estimated by the model, and whether it is a violation $g_o.violation$, as defined below.

We evaluate every subgroup $g\subseteq P$ of two or more points using our contextual model. 
We retain $g$ as a candidate if it is not a desired group, and if the model labels it positive with uncertainty ${\ge}0.9$ \footnote{We chose this threshold via hyperparameter tuning using the training data.}.    
We found that this alone results in a large number of redundant co-linear groups.   
For example, consider a linear line with four points.  
All subsets of the line are also co-linear and will be labeled positive by the model.   
For this reason, we do not consider $g$ a data-induced group if it is co-linear and a strict subset of another co-linear group.  

\begin{example}
\Cref{fig:app} uses the points in \Cref{fig:redesign} (left) as input, and specifies the three categories as desired groups $G_D = \{AB, CD, EF\}$.   
The model identifies 9 data-induced groups that a user is likely to perceive, of which only 2 of the desired groups are included.   
The user can filter for groups that contain ``A'', and the sidebar shows the corresponding 4 groups.   
The groups with blue points correspond to desired groups, while those in red are potential violations.  
The user then decides there are too many violations and plans to change the x-axis permutation.

\end{example}

\begin{figure}[t]
    \centering
    \includegraphics[width=\columnwidth, trim={0cm 0.5cm 0cm 0cm}]{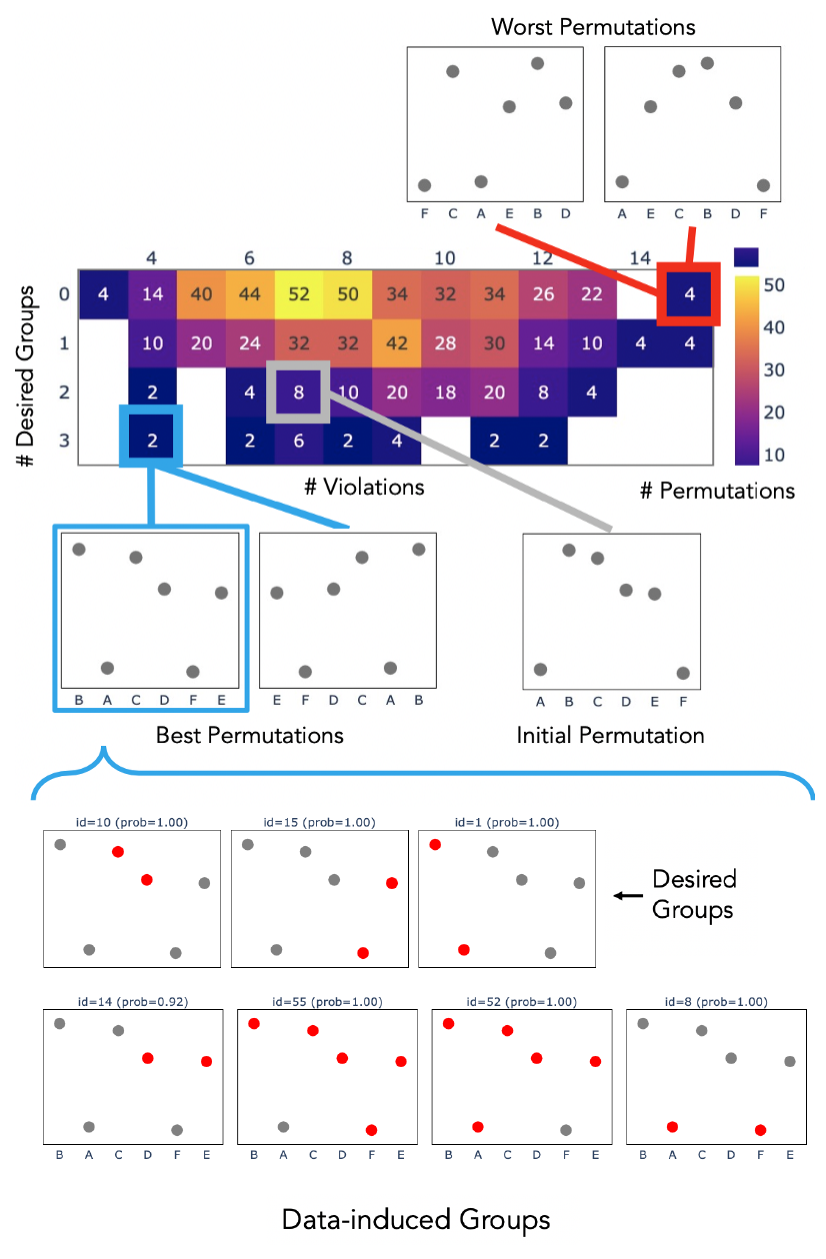}
    \vspace*{-.2in}
    
    \caption{There are 720 possible x-axis permutations for the chart in \cref{fig:redesign}.   The heatmap shows the distribution of perumatitons in terms of the number of violations (x-axis) and desired groups (y-axis) that a user is likely to see.  Each cell is labeled with the number of permutations.   We highlight two examples of the \blue{optimal design in blue} (all desired groups, 4 violations), the \red{worst designs in red} (no desired groups, 15 violations), and the \gray{initial permutation in gray} (2 desired groups, 7 violations).  The bottom lists all data-induced groups for the highlighted optimal design, \yilan{along with the trained model's predicted probability of those groupings.} }
    \label{fig:redesign3}
    \vspace*{-.2in}
\end{figure}

\subsection{Redesign Application}
Once a visualization reader forms visual groups, it becomes extremely effortful for them to break these groups to form new ones~\cite{bearfield2023does}.
This can lead to misinterpretation of data. 
As a result, it is important for designers to easily redesign their visualizations to reduce the number of violating groups.  
To this end, we extended the application to propose automatic redesigns that permute the x-axis in ways that ensure the desired groups are accentuated while breaking unintended groups  (\Cref{fig:app} (4-5)).  Once the designer has drawn their desired groups in the main chart (1), they can use a slider to specify whether to prioritize accentuating the desired groups or minimizing undesired groups (4).   They then press the ``Generate Redesign'' button, and the system solves an optimization problem to recommend the top permutations (5).   
The developer can click on any of the recommended permutations to update the sidebar with the list of data-induced groupings.

The optimization is defined as follows.
For a given permutation $P_i\in Permutations(P)$, we first evaluate $G_O = Diagnose(P_i, G_D)$ and then compute its score $s_i$ as the weighted sum of two factors.  $s_d(P_i)$ computes the total probability of desired groups in $G_D$ that were identified in $G_O$.  And $s_v(P_i)$ then penalizes the number of groups in $G_O$ that are violations:
\begin{align*}
s(P_i) &= \alpha \blue{s_d(P_i)} - (1-\alpha) \red{s_v(P_i)} \hspace{3em}\alpha\in[0,1]\\
\blue{s_d(P_i)} &= \sum_{g_o\in G_D\land g_o\in G_O} g_o.prob\hspace{2em} \\
\red{s_v(P_i)} &= |\{g_o\in g_O | g_o.violation\}|
\end{align*}

In addition, the designer can constrain the search space to a custom set of permutations.  
For instance, the data may contain a hierarchical structure $H$ that should be preserved in any recommended permutation: \Cref{fig:redesign} renders products by beverage category and the designer wants beverages in the same category to remain together in the permutation.  
Thus, a permutation that swaps, for example, B (a tea) and C (a coffee) would violate the hierarchy constraint.   In this case, permutations can only reorder points within a category (e.g., swap A and B) or reorder entire categories (e.g., swap tea and coffee).  This is specified by supplying a method $isValid(P)$ that must return true for a permutation to be valid.

Formally, we solve the following problem by exhaustively enumerating all permutations of the x-axis.
\begin{problem}[Redesign Problem]\label{p:redesign}
Given points $P$ and desired groups $G_D$, return $P^* = \argmax_{P_i\in Permutations(P)} s(P_i)$, where $isValid(P^*)$.
\end{problem}

%

%

\begin{example}
The designer wants the system to recommend a redesign, and uses the slider in \Cref{fig:app} (4) to specify that they prefer a permutation ensuring all desired groups in $G_D=\{AB,CD,EF\}$ will be perceived.
The user then clicks ``Generate Redesign.''  
The system evaluates all $6!=720$ permutations.   
The optimal permutations are listed (5) and the user can click on each to see the list of data-induced groups in the sidebar.   In this case, the optimal permutation has only 4 (instead of 7) violations and ensures all desired groups are seen. \yilan{This scenario highlights how a practitioner can iteratively use the tool to explore trade-offs, making explicit the otherwise implicit decisions about which visual biases to tolerate.}


\end{example}

To better understand the optimization landscape, \Cref{fig:redesign3} summarizes the distribution of all permutations in terms of the number of violations (x-axis) and number of desired groups that are met (y-axis). Notably, no permutation can guarantee three desired groups with no other data-induced groups (lower left corner).  There are only two permutations that guarantee all desired groups alongside 4 violations.   
In contrast, 356 permutations, or nearly $50\%$ of permutations, entail no desired group, and up to 15 violations in the worst case.

The \gray{initial permutation in grey} ABCDEF only shows two of the desired groups, and exhibits 7 violations.   
The two \blue{optimal permutations in blue} are mirrors of each other, and we list all the data-induced groups for the highlighted chart immediately below it.  We also list two examples of the \red{worst permutations in red}, and a cursory inspection shows false patterns (such as the parabolic shape) that are difficult to ignore.  
Ultimately, the large search space and prevalence of bad permutations motivate and justify adding an automated redesign and diagnosis tool to the designer's toolbox.

\section{Discussion, Limitations, and Future Directions}

Visualization enhances data interpretation but can lead to incorrect assumptions due to limitations in our perceptual systems~\cite{ceja2020truth,franconeri2021science}.
For example, Correll et al. have demonstrated that when participants are shown a line chart depicting how a value changes over 12 months and asked to find the month with the highest average value, they intuitively selected the month with the highest value rather than the month with the highest \textit{average} value~\cite{correll2012comparing}.
Such discrepancies arise when intuitive perceptions conflict with accurate data interpretation.
To address this, researchers have developed design guidelines based on empirical studies. Our work builds on this by examining illusory groups in dot plots and proposing automated solutions for improved visualization design. \yilan{From a usability standpoint, our workflow examples show how the tools integrate into common visualization tasks, allowing designers to both anticipate and iteratively refine grouping effects during chart creation.}



\stitle{Discussion.} 
\cindy{
Visualization readers make sense of visualizations by quickly segmenting and grouping marks according to classic Gestalt principles—proximity, similarity, good continuation and collinearity, common region, as well as connectedness.
This process organizes the visualization into perceived units for comparison~\cite{wertheimer1938, wagemans2012a, wagemans2012b, palmer1992, palmerrock1994, field1993}. 
Much of this organization occurs early and automatically, guided by pre-attentive mechanisms that steer attention toward structured units~\cite{treisman1980, wolfe2004}. These initial groupings then act as attentional anchors: once marks are bound into units, attention and comparison operate over those units, and therefore, breaking up already-formed groups to make new groups can be extremely effortful~\cite{egly1994, scholl2001, bearfield2023does}. 
Building on this foundation, our model operationalizes grouping-relevant features—spatial proximity, group size, collinearity, and common region to predict which data points are likely to be grouped. 
We demonstrate the dangers of arranging the data points in ways that accidentally emphasize illusory groups, and provide redesign recommendations to facilitate meaningful visual grouping.
}

Currently, visualization design maps data attributes to visual features, but {\it table contents} rarely influence this process.
Our work shows that visualization design alone is not sufficient to understand how users will perceive patterns in the visualization.   
Instead, the data values and distributions play a crucial role, and can sometimes overpower the user's logical understanding of the data (e.g., the data types).   
This implies that although a visualization design $V_1$ may in general be more effective than $V_2$, the latter may be more suitable for a specific dataset.   In other words, the properties of a dataset can necessitate changing the visualization design altogether.


\stitle{Limitations and Future Work.}
 First, our model makes predictions about each group of points in isolation, and implicitly accounts for points outside of the group in some of the clustering and separation features.   
However, data points in a visualization can interact with each other beyond clustering and separation to exert compounding effects on perception. 
For example, perceived data shape, such as symmetry and orthogonality can impact perception and memory~\cite{marriott2012memorability}, and
extreme values in a line chart can influence the perceived overall line shape and distort reader perception of the line's average position~\cite{savalia2022data}.
An important direction for future work is to incorporate additional perceptual interactions between data groups and points in the predictive model. \yilan{Moreover, our feature definitions are instantiated for dot plots and assume a reorderable categorical axis; extending analogous features to other chart families (e.g. bars and lines charts) and supporting fixed orderings is left to future work.}

Second, our model was trained on user-identified groups but synthetically generated negative examples. 
The generation process followed a simple heuristic to identify subsets of points that were ``obviously'' non-groups.   
However, this heuristic potentially introduces bias in the training dataset.  
Future work can explore other heuristics and recruit more users to identify  negative examples and create a larger dataset. Moreover, individual differences, such as spatial abilities, may influence data pattern detection and could be explored in future work~\cite{green2010towards}.
Additionally, the current work took a prediction modeling approach. 
Future work can broaden the scope by identifying people's grouping tendencies through qualitative feedback (e.g., think-aloud studies), and systematically testing and verifying the conditions under which certain grouping strategies are used.
For example, researchers can leverage large language models to generate visualizations under various hypotheses of grouping behaviors~\cite{tian2024chartgpt} and evaluate human responses to these visualizations~\cite{wang2024aligned}.

Third, our work focused on dot plots with exactly 6 points.  Although our model is designed to be independent of the number of points in a group and we suspect our findings generalize, more work is needed to evaluate conditions with more marks, other data types, visual encodings, and designs.  
For instance, all of our charts mapped the nominal attribute to the x-axis.  
Since existing work has illustrated potential asymmetry between how people perceive horizontal and vertical planes (e.g., a vertical line adjacent to a horizontal line of identical length is perceived as longer~\cite{robinson1998illusion}), future work can test out how our findings generalizes to cases with nominal y-axis or multi-dimensional planes. Moreover, using more than six points in the user studies to construct a training dataset could be valuable for testing the generalizability of our results in future research. 
\cindy{Further, our user studies explicitly instructed participants to search for groupings, which may only reflect a limited set of real-world visualization tasks. 
While this framing allowed us to isolate and examine grouping behavior in a controlled setting, future work should investigate how data-induced groupings emerge when users pursue more open-ended or domain-specific tasks.}

Moreover, our work highlights a gap in current visualization research regarding the intersection of spatial configurations and user interpretations in unconventional plots, such as dot plots without explicit gridlines. Future research could explore whether similar grouping behaviors occur in more complex visualizations, such as bar charts with overlapping categorical values, or in tasks where users are not explicitly prompted to identify groups. Furthermore, integrating concepts from visual separability metrics, such as those proposed by \cite{aupetit2016sepme}, with our predictive modeling framework could offer a more comprehensive approach to understanding and mitigating perceptual biases.


\section{Conclusions}

We conducted two user studies to understand the effects of {\it data-induced groupings}: perceptual groupings based on the values in the underlying dataset rather than the visual specification.   
The first study asked users to identify groups in dot plots while the second reminded users that the data is nominal.  
These studies reveal two key findings: (1) Co-linearity is a strong determinant of visual grouping, often overriding explicit instructions about the nominal nature of the data; and (2) Clustering features, such as proximity and centroid-based grouping, dominate when co-linearity cues are weak or absent.
We developed accurate models and showed how they can potentially be applied to aid visualization design.

\section*{Acknowledgments}
   This work was supported in part by the National Science Foundation awards IIS-2237585, IIS-2311575, III-2453462, IIS-2107490, IIS-1901485, 1845638, 1740305, 2008295, 2106197, 2103794, 2312991, and by Amazon, Google, Adobe, the Center for Artificial Intelligence and Technology (CAIT), Columbia SIRS, and Intellect Design.

\bibliography{references}

@incollection{wertheimer1938,
  author    = {Wertheimer, Max},
  title     = {Laws of Organization in Perceptual Forms},
  booktitle = {A Source Book of Gestalt Psychology},
  editor    = {Ellis, Willis D.},
  year      = {1938},
  publisher = {Routledge \& Kegan Paul},
  address   = {London},
  pages     = {71--88},
  note      = {Original work published 1923}
}

@article{wagemans2012a,
  author  = {Wagemans, Johan and Elder, James H. and Kubovy, Michael and Palmer, Stephen E. and Peterson, Mary A. and Singh, Manish and von der Heydt, R{\"u}diger},
  title   = {A century of Gestalt psychology in visual perception: I. Perceptual grouping and figure–ground organization},
  journal = {Psychological Bulletin},
  year    = {2012},
  volume  = {138},
  number  = {6},
  pages   = {1172--1217},
  doi     = {10.1037/a0029333}
}

@article{holder2022dispersion,
  title={Dispersion vs disparity: Hiding variability can encourage stereotyping when visualizing social outcomes},
  author={Holder, Eli and Xiong, Cindy},
  journal={IEEE Transactions on Visualization and Computer Graphics},
  volume={29},
  number={1},
  pages={624--634},
  year={2022},
  publisher={IEEE}
}

@incollection{rensink2013prospects,
  title={On the prospects for a science of visualization},
  author={Rensink, Ronald A},
  booktitle={Handbook of human centric visualization},
  pages={147--175},
  year={2013},
  publisher={Springer}
}

@article{newman2012bar,
  title={Bar graphs depicting averages are perceptually misinterpreted: The within-the-bar bias},
  author={Newman, George E and Scholl, Brian J},
  journal={Psychonomic bulletin \& review},
  volume={19},
  number={4},
  pages={601--607},
  year={2012},
  publisher={Springer}
}

@article{kerns2021two,
  title={Two graphs walk into a bar: Readout-based measurement reveals the Bar-Tip Limit error, a common, categorical misinterpretation of mean bar graphs},
  author={Kerns, Sarah H and Wilmer, Jeremy B},
  journal={Journal of vision},
  volume={21},
  number={12},
  pages={17--17},
  year={2021},
  publisher={The Association for Research in Vision and Ophthalmology}
}

@article{wagemans2012b,
  author  = {Wagemans, Johan and Feldman, Jacob and Gepshtein, Sergei and Kimchi, Ruth and Pomerantz, James R. and van der Helm, Peter A. and van Leeuwen, Cees},
  title   = {A century of Gestalt psychology in visual perception: II. Conceptual and theoretical foundations},
  journal = {Psychological Bulletin},
  year    = {2012},
  volume  = {138},
  number  = {6},
  pages   = {1218--1252},
  doi     = {10.1037/a0029334}
}

@article{palmer1992,
  author  = {Palmer, Stephen E.},
  title   = {Common region: A new principle of perceptual grouping},
  journal = {Cognitive Psychology},
  year    = {1992},
  volume  = {24},
  number  = {3},
  pages   = {436--447},
  doi     = {10.1016/0010-0285(92)90014-S}
}

@article{palmerrock1994,
  author  = {Palmer, Stephen E. and Rock, Irvin},
  title   = {Rethinking perceptual organization: The role of uniform connectedness},
  journal = {Psychological Review},
  year    = {1994},
  volume  = {101},
  number  = {4},
  pages   = {725--753},
  doi     = {10.1037/0033-295X.101.4.725}
}

@article{field1993,
  author  = {Field, David J. and Hayes, Andrew and Hess, Robert F.},
  title   = {Contour integration by the human visual system: Evidence for a local ``association field''},
  journal = {Vision Research},
  year    = {1993},
  volume  = {33},
  number  = {2},
  pages   = {173--193},
  doi     = {10.1016/0042-6989(93)90156-Q}
}

@article{treisman1980,
  author  = {Treisman, Anne M. and Gelade, Garry},
  title   = {A feature-integration theory of attention},
  journal = {Cognitive Psychology},
  year    = {1980},
  volume  = {12},
  number  = {1},
  pages   = {97--136},
  doi     = {10.1016/0010-0285(80)90005-5}
}

@article{wolfe2004,
  author  = {Wolfe, Jeremy M. and Horowitz, Todd S.},
  title   = {What attributes guide the deployment of visual attention and how do they do it?},
  journal = {Nature Reviews Neuroscience},
  year    = {2004},
  volume  = {5},
  number  = {6},
  pages   = {495--501},
  doi     = {10.1038/nrn1411}
}

@article{egly1994,
  author  = {Egly, R. and Driver, J. and Rafal, R. D.},
  title   = {Shifting visual attention between objects and locations: Evidence from normal and parietal lesion subjects},
  journal = {Journal of Experimental Psychology: General},
  year    = {1994},
  volume  = {123},
  number  = {2},
  pages   = {161--177},
  doi     = {10.1037/0096-3445.123.2.161}
}

@article{scholl2001,
  author  = {Scholl, Brian J.},
  title   = {Objects and attention: The state of the art},
  journal = {Cognition},
  year    = {2001},
  volume  = {80},
  number  = {1-2},
  pages   = {1--46},
  doi     = {10.1016/S0010-0277(00)00152-9}
}

@inproceedings{green2010towards,
  title={Towards the personal equation of interaction: The impact of personality factors on visual analytics interface interaction},
  author={Green, Tear Marie and Fisher, Brian},
  booktitle={2010 IEEE Symposium on Visual Analytics Science and Technology},
  pages={203--210},
  year={2010},
  organization={IEEE}
}

@inproceedings{borgo2018information,
  title={Information visualization evaluation using crowdsourcing},
  author={Borgo, Rita and Micallef, Luana and Bach, Benjamin and McGee, Fintan and Lee, Bongshin},
  booktitle={Computer Graphics Forum},
  volume={37},
  number={3},
  pages={573--595},
  year={2018},
  organization={Wiley Online Library}
}

@inproceedings{borgo2017crowdsourcing,
  title={Crowdsourcing for information visualization: Promises and pitfalls},
  author={Borgo, Rita and Lee, Bongshin and Bach, Benjamin and Fabrikant, Sara and Jianu, Radu and Kerren, Andreas and Kobourov, Stephen and McGee, Fintan and Micallef, Luana and von Landesberger, Tatiana and others},
  booktitle={Evaluation in the Crowd. Crowdsourcing and Human-Centered Experiments: Dagstuhl Seminar 15481, Dagstuhl Castle, Germany, November 22--27, 2015, Revised Contributions},
  pages={96--138},
  year={2017},
  organization={Springer}
}

@inproceedings{aupetit2016sepme,
  title={Sepme: 2002 new visual separation measures},
  author={Aupetit, Micha{\"e}l and Sedlmair, Michael},
  booktitle={2016 IEEE pacific visualization symposium (PacificVis)},
  pages={1--8},
  year={2016},
  organization={IEEE}
}

@inproceedings{zhao2019neighborhood,
  title={Neighborhood perception in bar charts},
  author={Zhao, Mingqian and Qu, Huamin and Sedlmair, Michael},
  booktitle={Proceedings of the 2019 CHI Conference on Human Factors in Computing Systems},
  pages={1--12},
  year={2019}
}

@article{marriott2012memorability,
  title={Memorability of visual features in network diagrams},
  author={Marriott, Kim and Purchase, Helen and Wybrow, Michael and Goncu, Cagatay},
  journal={IEEE Transactions on Visualization and Computer Graphics},
  volume={18},
  number={12},
  pages={2477--2485},
  year={2012},
  publisher={IEEE}
}

@article{yantis1992multielement,
  title={Multielement visual tracking: Attention and perceptual organization},
  author={Yantis, Steven},
  journal={Cognitive psychology},
  volume={24},
  number={3},
  pages={295--340},
  year={1992},
  publisher={Elsevier}
}

@book{robinson1998illusion,
  title={{The Psychology of Visual Illusion}},
  author={Robinson, J O},
  year={1998},
  publisher={Courier Corp}
}

@article{treisman1982perceptual,
  title={Perceptual grouping and attention in visual search for features and for objects.},
  author={Treisman, Anne},
  journal={Journal of experimental psychology: human perception and performance},
  volume={8},
  number={2},
  pages={194},
  year={1982},
  publisher={American Psychological Association}
}

@article{savalia2022data,
  title={Data Shape and Response Modalities Can Bias Estimations of Average Data Location in Visualizations},
  author={Savalia, Tejas and Ceja, Cristina and Cowell, Rosemary and Xiong, Cindy},
  journal={Journal of Vision},
  volume={22},
  number={14},
  pages={3413--3413},
  year={2022},
  publisher={The Association for Research in Vision and Ophthalmology}
}

@article{ceja2020truth,
  title={Truth or square: Aspect ratio biases recall of position encodings},
  author={Ceja, Cristina R and McColeman, Caitlyn M and Xiong, Cindy and Franconeri, Steven L},
  journal={IEEE Transactions on Visualization and Computer Graphics},
  volume={27},
  number={2},
  pages={1054--1062},
  year={2020},
  publisher={IEEE}
}

@inproceedings{correll2012comparing,
  title={Comparing averages in time series data},
  author={Correll, Michael and Albers, Danielle and Franconeri, Steven and Gleicher, Michael},
  booktitle={Proceedings of the SIGCHI Conference on Human Factors in Computing Systems},
  pages={1095--1104},
  year={2012}
}

@article{lundberg2017unified,
	title        = {A unified approach to interpreting model predictions},
	author       = {Lundberg, Scott M and Lee, Su-In},
	year         = 2017,
	journal      = {Advances in neural information processing systems},
	volume       = 30
}

@inproceedings{kim2018assessing,
  title={Assessing effects of task and data distribution on the effectiveness of visual encodings},
  author={Kim, Younghoon and Heer, Jeffrey},
  booktitle={Computer Graphics Forum},
  volume={37},
  number={3},
  pages={157--167},
  year={2018},
  organization={Wiley Online Library}
}

@article{heer2006multi,
	title        = {Multi-scale banking to 45 degrees},
	author       = {Heer, Jeffrey and Agrawala, Maneesh},
	year         = 2006,
	journal      = {IEEE Transactions on Visualization and Computer Graphics},
	publisher    = {IEEE},
	volume       = 12,
	number       = 5,
	pages        = {701--708}
}

@article{bearfield2023does,
	title        = {What does the chart say? grouping cues guide viewer comparisons and conclusions in bar charts},
	author       = {Bearfield, Cindy Xiong and Stokes, Chase and Lovett, Andrew and Franconeri, Steven},
	year         = 2023,
	journal      = {IEEE Transactions on Visualization and Computer Graphics},
	publisher    = {IEEE}
}

@article{wang2017there,
	title        = {Is there a robust technique for selecting aspect ratios in line charts?},
	author       = {Wang, Yunhai and Wang, Zeyu and Zhu, Lifeng and Zhang, Jian and Fu, Chi-Wing and Cheng, Zhanglin and Tu, Changhe and Chen, Baoquan},
	year         = 2017,
	journal      = {IEEE Transactions on Visualization and Computer Graphics},
	publisher    = {IEEE},
	volume       = 24,
	number       = 12,
	pages        = {3096--3110}
}

@article{palan2018prolific,
	title        = {Prolific. ac—A subject pool for online experiments},
	author       = {Palan, Stefan and Schitter, Christian},
	year         = 2018,
	journal      = {Journal of Behavioral and Experimental Finance},
	publisher    = {Elsevier},
	volume       = 17,
	pages        = {22--27}
}

@article{kindlmann2014algebraic,
	title        = {An algebraic process for visualization design},
	author       = {Kindlmann, Gordon and Scheidegger, Carlos},
	year         = 2014,
	journal      = {IEEE transactions on visualization and computer graphics},
	publisher    = {IEEE},
	volume       = 20,
	number       = 12,
	pages        = {2181--2190}
}

@article{xiong2019illusion,
	title        = {Illusion of causality in visualized data},
	author       = {Xiong, Cindy and Shapiro, Joel and Hullman, Jessica and Franconeri, Steven},
	year         = 2019,
	journal      = {IEEE transactions on visualization and computer graphics},
	publisher    = {IEEE},
	volume       = 26,
	number       = 1,
	pages        = {853--862}
}

@article{cleveland1984graphical,
	title        = {Graphical perception: Theory, experimentation, and application to the development of graphical methods},
	author       = {Cleveland, William S and McGill, Robert},
	year         = 1984,
	journal      = {Journal of the American statistical association},
	publisher    = {Taylor \& Francis Group},
	volume       = 79,
	number       = 387,
	pages        = {531--554}
}

@inproceedings{amar2004,
	title        = {A Knowledge Task-Based Framework for Design and Evaluation of Information Visualizations},
	author       = {R. {Amar} and J. {Stasko}},
	year         = 2004,
	booktitle    = {IEEE Symposium on Information Visualization},
	volume       = {},
	number       = {},
	pages        = {143--150}
}

@article{franconeri2021science,
	title        = {The science of visual data communication: What works},
	author       = {Franconeri, S and Padilla, L and Shah, P and Zacks, J and Hullman, J},
	year         = 2021,
	journal      = {Psychological Science in the Public Interest},
	publisher    = {SAGE Publications Sage CA: Los Angeles, CA},
	volume       = 22,
	number       = 3,
	pages        = {110--161}
}

@article{lee2016vlat,
	title        = {Vlat: Development of a visualization literacy assessment test},
	author       = {Lee, Sukwon and Kim, Sung-Hee and Kwon, Bum Chul},
	year         = 2016,
	journal      = {IEEE transactions on visualization and computer graphics},
	publisher    = {IEEE},
	volume       = 23,
	number       = 1,
	pages        = {551--560}
}

@article{xiong2021visual,
	title        = {Visual arrangements of bar charts influence comparisons in viewer takeaways},
	author       = {Xiong, Cindy and Setlur, Vidya and Bach, Benjamin and Lin, Kylie and Koh, Eunyee and Franconeri, Steven},
	year         = 2021,
	journal      = {IEEE Transactions on Visualization and Computer Graphics},
	publisher    = {IEEE},
	volume       = 28,
	number       = 1,
	pages        = {955--965}
}

@article{nothelfer2019measures,
	title        = {Measures of the benefit of direct encoding of data deltas for data pair relation perception},
	author       = {Nothelfer, Christine and Franconeri, Steven},
	year         = 2019,
	journal      = {IEEE transactions on visualization and computer graphics},
	publisher    = {IEEE},
	volume       = 26,
	number       = 1,
	pages        = {311--320}
}

@article{shah2011bar,
	title        = {Bar and line graph comprehension: An interaction of top-down and bottom-up processes},
	author       = {Shah, Priti and Freedman, Eric G},
	year         = 2011,
	journal      = {Topics in cognitive science},
	publisher    = {Wiley Online Library},
	volume       = 3,
	number       = 3,
	pages        = {560--578}
}

@article{mackinlay1986automating,
	title        = {Automating the design of graphical presentations of relational information},
	author       = {Mackinlay, Jock},
	year         = 1986,
	journal      = {Acm Transactions On Graphics (Tog)},
	publisher    = {Acm New York, NY, USA},
	volume       = 5,
	number       = 2,
	pages        = {110--141}
}

@misc{adobe,
	title        = {Adobe Analytics},
	year         = 2020,
	howpublished = {\url{https://business.adobe.com/products/marketing-cloud/main.html}},
	key          = {Adobe}
}

@article{zacks1999bars,
	title        = {Bars and lines: A study of graphic communication},
	author       = {Zacks, Jeff and Tversky, Barbara},
	year         = 1999,
	journal      = {Memory \& cognition},
	publisher    = {Springer},
	volume       = 27,
	number       = 6,
	pages        = {1073--1079}
}

@article{saket2018task,
	title        = {Task-based effectiveness of basic visualizations},
	author       = {Saket, Bahador and Endert, Alex and Demiralp, Cagatay},
	year         = 2018,
	journal      = {IEEE transactions on visualization and computer graphics},
	publisher    = {IEEE},
	volume       = 25,
	number       = 7,
	pages        = {2505--2512}
}

@article{xiong2021vss,
	title        = {Visual salience and grouping cues guide relation perception in visual data displays},
	author       = {Xiong, Cindy and Stokes, Chase and Franconeri, Steve},
	year         = 2021,
	journal      = {Journal of Vision},
	publisher    = {The Association for Research in Vision and Ophthalmology},
	volume       = 21,
	number       = 9,
	pages        = {2095--2095}
}

@article{chen2020co,
	title        = {Co-bridges: Pair-wise visual connection and comparison for multi-item data streams},
	author       = {Chen, Siming and Andrienko, Natalia and Andrienko, Gennady and Li, Jie and Yuan, Xiaoru},
	year         = 2020,
	journal      = {IEEE Transactions on Visualization and Computer Graphics},
	publisher    = {IEEE},
	volume       = 27,
	number       = 2,
	pages        = {1612--1622}
}

@article{mccoleman2021rethinking,
	title        = {Rethinking the ranks of visual channels},
	author       = {McColeman, Caitlyn M and Yang, Fumeng and Brady, Timothy F and Franconeri, Steven},
	year         = 2021,
	journal      = {IEEE Transactions on Visualization and Computer Graphics},
	publisher    = {IEEE},
	volume       = 28,
	number       = 1,
	pages        = {707--717}
}

@article{joos2022visual,
	title        = {Visual Comparison of Networks in VR},
	author       = {Joos, Lucas and Jaeger-Honz, Sabrina and Schreiber, Falk and Keim, Daniel A and Klein, Karsten},
	year         = 2022,
	journal      = {IEEE Transactions on Visualization and Computer Graphics},
	publisher    = {IEEE},
	volume       = 28,
	number       = 11,
	pages        = {3651--3661}
}

@inproceedings{pister2023combinet,
	title        = {ComBiNet: Visual Query and Comparison of Bipartite Multivariate Dynamic Social Networks},
	author       = {Pister, Alexis and Prieur, Christophe and Fekete, Jean-Daniel},
	year         = 2023,
	booktitle    = {Computer Graphics Forum}
}

@article{lyi2020comparative,
	title        = {Comparative layouts revisited: Design space, guidelines, and future directions},
	author       = {LYi, Sehi and Jo, Jaemin and Seo, Jinwook},
	year         = 2020,
	journal      = {IEEE Transactions on Visualization and Computer Graphics},
	publisher    = {IEEE},
	volume       = 27,
	number       = 2,
	pages        = {1525--1535}
}

@inproceedings{battle2019characterizing,
	title        = {Characterizing exploratory visual analysis: A literature review and evaluation of analytic provenance in tableau},
	author       = {Battle, Leilani and Heer, Jeffrey},
	year         = 2019,
	booktitle    = {Computer graphics forum},
	volume       = 38,
	number       = 3,
	pages        = {145--159},
	organization = {Wiley Online Library}
}

@inproceedings{rae2022understanding,
	title        = {Understanding Visual Investigation Patterns Through Digital “Field” Observations},
	author       = {Rae, Irene and Zhou, Feng and Bilsing, Martin and Bunge, Philipp},
	year         = 2022,
	booktitle    = {Proceedings of the 2022 CHI Conference on Human Factors in Computing Systems},
	pages        = {1--16}
}

@inproceedings{pandey2016towards,
	title        = {Towards understanding human similarity perception in the analysis of large sets of scatter plots},
	author       = {Pandey, Anshul Vikram and Krause, Josua and Felix, Cristian and Boy, Jeremy and Bertini, Enrico},
	year         = 2016,
	booktitle    = {Proceedings of the 2016 CHI Conference on Human Factors in Computing Systems},
	pages        = {3659--3669}
}

@article{wu2022view,
	title        = {View composition algebra for ad hoc comparison},
	author       = {Wu, Eugene},
	year         = 2022,
	journal      = {IEEE Transactions on Visualization and Computer Graphics},
	publisher    = {IEEE},
	volume       = 28,
	number       = 6,
	pages        = {2470--2485}
}

@article{gaba2022comparison,
	title        = {Comparison conundrum and the chamber of visualizations: An exploration of how language influences visual design},
	author       = {Gaba, Aimen and Setlur, Vidya and Srinivasan, Arjun and Hoffswell, Jane and Xiong, Cindy},
	year         = 2022,
	journal      = {IEEE Transactions on Visualization and Computer Graphics},
	publisher    = {IEEE},
	volume       = 29,
	number       = 1,
	pages        = {1211--1221}
}

@inproceedings{hearst2019toward,
	title        = {Toward Interface Defaults for Vague Modifiers in Natural Language Interfaces for Visual Analysis},
	author       = {Hearst, Marti and Tory, Melanie and Setlur, Vidya},
	year         = 2019,
	booktitle    = {2019 IEEE Visualization Conference (VIS)},
	volume       = {},
	number       = {},
	pages        = {21--25},
	doi          = {10.1109/VISUAL.2019.8933569}
}

@article{lovett2017modeling,
	title        = {Modeling visual problem solving as analogical reasoning.},
	author       = {Lovett, Andrew and Forbus, Kenneth},
	year         = 2017,
	journal      = {Psychological review},
	publisher    = {American Psychological Association},
	volume       = 124,
	number       = 1,
	pages        = 60
}

@article{love1999structural,
	title        = {A structural account of global and local processing},
	author       = {Love, Bradley C and Rouder, Jeffrey N and Wisniewski, Edward J},
	year         = 1999,
	journal      = {Cognitive psychology},
	publisher    = {Elsevier},
	volume       = 38,
	number       = 2,
	pages        = {291--316}
}

@article{navon1977forest,
	title        = {Forest before trees: The precedence of global features in visual perception},
	author       = {Navon, David},
	year         = 1977,
	journal      = {Cognitive psychology},
	publisher    = {Elsevier},
	volume       = 9,
	number       = 3,
	pages        = {353--383}
}

@article{ahissar2004reverse,
	title        = {The reverse hierarchy theory of visual perceptual learning},
	author       = {Ahissar, Merav and Hochstein, Shaul},
	year         = 2004,
	journal      = {Trends in cognitive sciences},
	publisher    = {Elsevier},
	volume       = 8,
	number       = 10,
	pages        = {457--464}
}

@article{michal2016visual,
	title        = {Visual routines for extracting magnitude relations},
	author       = {Michal, Audrey L and Uttal, David and Shah, Priti and Franconeri, Steven L},
	year         = 2016,
	journal      = {Psychonomic bulletin \& review},
	publisher    = {Springer},
	volume       = 23,
	number       = 6,
	pages        = {1802--1809}
}

@article{jardine2019perceptual,
	title        = {The Perceptual Proxies of Visual Comparison},
	author       = {Jardine, Nicole and Ondov, Brian D and Elmqvist, Niklas and Franconeri, Steven},
	year         = 2019,
	journal      = {IEEE transactions on visualization and computer graphics},
	publisher    = {IEEE},
	volume       = 26,
	number       = 1,
	pages        = {1012--1021}
}

@article{duncan1989visual,
	title        = {Visual search and stimulus similarity.},
	author       = {Duncan, John and Humphreys, Glyn W},
	year         = 1989,
	journal      = {Psychological review},
	publisher    = {American Psychological Association},
	volume       = 96,
	number       = 3,
	pages        = 433
}

@article{burlinson2017open,
	title        = {Open vs. Closed Shapes: New Perceptual Categories?},
	author       = {Burlinson, David and Subramanian, Kalpathi and Goolkasian, Paula},
	year         = 2017,
	journal      = {IEEE transactions on visualization and computer graphics},
	publisher    = {IEEE},
	volume       = 24,
	number       = 1,
	pages        = {574--583}
}

@article{yu2019similarity,
	title        = {Similarity grouping as feature-based selection},
	author       = {Yu, Dian and Xiao, Xiao and Bemis, Douglas K and Franconeri, Steven L},
	year         = 2019,
	journal      = {Psychological science},
	publisher    = {Sage Publications Sage CA: Los Angeles, CA},
	volume       = 30,
	number       = 3,
	pages        = {376--385}
}

@article{wang2024aligned,
  title={How Aligned are Human Chart Takeaways and LLM Predictions? A Case Study on Bar Charts with Varying Layouts},
  author={Wang, Huichen Will and Hoffswell, Jane and Bursztyn, Victor S and Bearfield, Cindy Xiong and others},
  journal={IEEE Transactions on Visualization and Computer Graphics},
  year={2024},
  publisher={IEEE}
}

@article{tian2024chartgpt,
  title={Chartgpt: Leveraging llms to generate charts from abstract natural language},
  author={Tian, Yuan and Cui, Weiwei and Deng, Dazhen and Yi, Xinjing and Yang, Yurun and Zhang, Haidong and Wu, Yingcai},
  journal={IEEE Transactions on Visualization and Computer Graphics},
  year={2024},
  publisher={IEEE}
}

@article{wagemans2012century,
	title        = {A century of Gestalt psychology in visual perception: I. Perceptual grouping and figure--ground organization.},
	author       = {Wagemans, Johan and Elder, James H and Kubovy, Michael and Palmer, Stephen E and Peterson, Mary A and Singh, Manish and von der Heydt, R{\"u}diger},
	year         = 2012,
	journal      = {Psychological bulletin},
	publisher    = {American Psychological Association},
	volume       = 138,
	number       = 6,
	pages        = 1172
}

@article{gleicher2013perception,
	title        = {Perception of average value in multiclass scatterplots},
	author       = {Gleicher, Michael and Correll, Michael and Nothelfer, Christine and Franconeri, Steven},
	year         = 2013,
	journal      = {IEEE transactions on visualization and computer graphics},
	publisher    = {IEEE},
	volume       = 19,
	number       = 12,
	pages        = {2316--2325}
}

@article{yu2019gestalt,
	title        = {Gestalt similarity groupings are not constructed in parallel},
	author       = {Yu, Dian and Tam, Derek and Franconeri, Steven L},
	year         = 2019,
	journal      = {Cognition},
	publisher    = {Elsevier},
	volume       = 182,
	pages        = {8--13}
}

@article{yu2014grouping,
  title={Grouping by similarity is serial, irrespective of spacing or group size},
  author={Yu, Dian and Tam, Derek and Franconeri, Steven},
  journal={Journal of Vision},
  volume={14},
  number={10},
  pages={808--808},
  year={2014},
  publisher={The Association for Research in Vision and Ophthalmology}
}

@article{franconeri2009number,
	title        = {Number estimation relies on a set of segmented objects},
	author       = {Franconeri, Steven L and Bemis, Douglas K and Alvarez, George Angelo},
	year         = 2009,
	journal      = {Cognition},
	publisher    = {Elsevier},
	volume       = 113,
	number       = 1,
	pages        = {1--13}
}

@article{brooks2015traditional,
	title        = {Traditional and new principles of perceptual grouping},
	author       = {Brooks, Joseph L},
	year         = 2015,
	publisher    = {Oxford Handbook of Perceptual Organization: Oxford University Press}
}

@article{ondov2018face,
	title        = {Face to face: Evaluating visual comparison},
	author       = {Ondov, Brian and Jardine, Nicole and Elmqvist, Niklas and Franconeri, Steven},
	year         = 2018,
	journal      = {IEEE transactions on visualization and computer graphics},
	publisher    = {IEEE},
	volume       = 25,
	number       = 1,
	pages        = {861--871}
}

@article{pinker1990theory,
	title        = {A theory of graph comprehension},
	author       = {Pinker, Steven},
	year         = 1990,
	journal      = {Artificial intelligence and the future of testing},
	pages        = {73--126}
}

@article{shah1999graphs,
	title        = {Graphs as aids to knowledge construction: Signaling techniques for guiding the process of graph comprehension.},
	author       = {Shah, Priti and Mayer, Richard E and Hegarty, Mary},
	year         = 1999,
	journal      = {Journal of educational psychology},
	publisher    = {American Psychological Association},
	volume       = 91,
	number       = 4,
	pages        = 690
}

@article{michal2017visual,
	title        = {Visual routines are associated with specific graph interpretations},
	author       = {Michal, Audrey L and Franconeri, Steven L},
	year         = 2017,
	journal      = {Cognitive Research: Principles and Implications},
	publisher    = {SpringerOpen},
	volume       = 2,
	number       = 1,
	pages        = {1--10}
}

@article{franconeri2013nature,
	title        = {The nature and status of visual resources.},
	author       = {Franconeri, Steven L},
	year         = 2013,
	publisher    = {Oxford University Press}
}

@article{Gleicher2011,
	title        = {Visual Comparison for Information Visualization},
	author       = {Gleicher, Michael and Albers, Danielle and Walker, Rick and Jusufi, Ilir and Hansen, Charles D. and Roberts, Jonathan C.},
	year         = 2011,
	month        = oct,
	journal      = {Information Visualization},
	publisher    = {Palgrave Macmillan},
	volume       = 10,
	number       = 4,
	pages        = {289–309},
	doi          = {10.1177/1473871611416549},
	issn         = {1473-8716},
	url          = {https://doi.org/10.1177/1473871611416549},
	issue_date   = {October 2011},
	numpages     = 21
}

@article{best2007perception,
  title={Perception of linear and nonlinear trends: using slope and curvature information to make trend discriminations},
  author={Best, Lisa A and Smith, Laurence D and Stubbs, D Alan},
  journal={Perceptual and motor skills},
  volume={104},
  number={3},
  pages={707--721},
  year={2007},
  publisher={SAGE Publications Sage CA: Los Angeles, CA}
}

@article{yang2018correlation,
  title={Correlation judgment and visualization features: A comparative study},
  author={Yang, Fumeng and Harrison, Lane T and Rensink, Ronald A and Franconeri, Steven L and Chang, Remco},
  journal={IEEE transactions on visualization and computer graphics},
  volume={25},
  number={3},
  pages={1474--1488},
  year={2018},
  publisher={IEEE}
}

@article{quadri2021survey,
  title={A survey of perception-based visualization studies by task},
  author={Quadri, Ghulam Jilani and Rosen, Paul},
  journal={IEEE transactions on visualization and computer graphics},
  volume={28},
  number={12},
  pages={5026--5048},
  year={2021},
  publisher={IEEE}
}

@article{larsen:1998,
	title        = {Effects of Spatial Separation in Visual Pattern Matching: Evidence on the Role of Mental Translation},
	author       = {Larsen, Axel and Bundesen, Claus},
	year         = 1998,
	month        = {07},
	journal      = {Journal of experimental psychology. Human perception and performance},
	volume       = 24,
	pages        = {719--31},
	doi          = {10.1037/0096-1523.24.3.719}
}

@article{posner1980,
	title        = {{Orienting of attention}},
	author       = {Posner, Michael I},
	year         = 1980,
	journal      = {The Quarterly Journal of Experimental Psychology},
	volume       = 32,
	number       = 1,
	pages        = {3--25},
	pmid         = 7367577
}

@article{folk2010,
	title        = {{Target-uncertainty effects in attentional capture: Color-singleton set or multiple attentional control settings?}},
	author       = {Folk, Charles L. and Anderson, Brian A.},
	year         = 2010,
	journal      = {Psychonomic Bulletin and Review},
	volume       = 17,
	number       = 3,
	pages        = {421--426}
}

@article{fygenson2023arrangement,
	title        = {The arrangement of marks impacts afforded messages: ordering, partitioning, spacing, and coloring in bar charts},
	author       = {Fygenson, Racquel and Franconeri, Steven and Bertini, Enrico},
	year         = 2023,
	journal      = {IEEE transactions on visualization and computer graphics},
	publisher    = {IEEE}
}

@article{franconeri2021three,
	title        = {Three perceptual tools for seeing and understanding visualized data},
	author       = {Franconeri, Steven L},
	year         = 2021,
	journal      = {Current Directions in Psychological Science},
	publisher    = {SAGE Publications Sage CA: Los Angeles, CA},
	volume       = 30,
	number       = 5,
	pages        = {367--375}
}

@article{kimchi1992primacy,
  title={Primacy of wholistic processing and global/local paradigm: a critical review.},
  author={Kimchi, Ruth},
  journal={Psychological bulletin},
  volume={112},
  number={1},
  pages={24},
  year={1992},
  publisher={American Psychological Association}
}

@article{palmer1977hierarchical,
  title={Hierarchical structure in perceptual representation},
  author={Palmer, Stephen E},
  journal={Cognitive psychology},
  volume={9},
  number={4},
  pages={441--474},
  year={1977},
  publisher={Elsevier}
}

@article{wertheimer1938laws,
  title={Laws of organization in perceptual forms.},
  author={Wertheimer, Max},
  year={1938},
  publisher={Kegan Paul, Trench, Trubner \& Company}
}

@article{prinzmetal1977good,
  title={Good continuation affects visual detection},
  author={Prinzmetal, William and Banks, William P},
  journal={Perception \& Psychophysics},
  volume={21},
  number={5},
  pages={389--395},
  year={1977},
  publisher={Springer}
}

@incollection{julesz1969cluster,
  title={Cluster formation at various perceptual levels},
  author={Julesz, Bela},
  booktitle={Methodologies of Pattern Recognition},
  pages={297--315},
  year={1969},
  publisher={Elsevier}
}

@article{kim1999grouping,
  title={Grouping effects on spatial attention in visual search},
  author={Kim, Min-Shik and Cave, Kyle R},
  journal={The Journal of general psychology},
  volume={126},
  number={4},
  pages={326--352},
  year={1999},
  publisher={Taylor \& Francis}
}

@article{gleicher2017considerations,
  title={Considerations for visualizing comparison},
  author={Gleicher, Michael},
  journal={IEEE Transactions on Visualization and Computer Graphics},
  volume={24},
  number={1},
  pages={413--423},
  year={2017},
  publisher={IEEE}
}
\bibliographystyle{IEEEtran}

\section{Biography Section}
 




\begin{IEEEbiographynophoto}{Yilan Jiang}
    received the MS in data science and currently pursuing the Ph.D. degree in Industrial and Enterprise Systems Engineering at the University of Illinois Urbana-Champaign. Her research mainly focusing on data-driven product design, optimization, and applications of machine learning. 
\end{IEEEbiographynophoto}
\begin{IEEEbiographynophoto}{Cindy Xiong Bearfield}
is an Assistant Professor in the School of Interactive Computing at the Georgia Institute of Technology. She received her Ph.D. in Cognitive Psychology and her MS in Statistics from Northwestern University. Bridging the fields of psychology and data visualization, she aims to understand the cognitive and perceptual processes that underlie visual data interpretation and communication. 
\end{IEEEbiographynophoto}
\begin{IEEEbiographynophoto}{Steven Franconeri}
is a Professor of Psychology at Northwestern University, and Director of the Northwestern Cognitive Science Program. He studies visual thinking and visual communication, across psychology, education, and information visualization.    
\end{IEEEbiographynophoto}
\begin{IEEEbiographynophoto}{Eugene Wu}
is an Associate Professor at
Columbia University, and Co-director of the Data, Agents, and Processes Lab (DAPLab). His research interests are
in systems for human data interaction. He has
made contributions across data systems, visualization, data cleaning and extraction, machine learning, explanation, crowd-sourcing, and agent systems. 
\end{IEEEbiographynophoto}

\vfill

\end{document}